\documentclass[useAMS,usenatbib]{mn2e}
\usepackage{epsfig}
\usepackage{times}
\usepackage{subeqn}
\usepackage{amssymb}

\newcommand{\abs}[1]{\left\vert#1\right\vert}

\def \Integers {\mathbb{Z}}

\def \dash {^{\,\prime}}

\def \grad {\nabla} \def \del {\nabla}
\def \centreline {\centerline}
    
\def \beqn {\begin{equation}}
\def \eeqn {\end{equation}}
\def \bdm {\begin{displaymath}}
\def \edm {\end{displaymath}}

\newcommand{\vect}[1]{\mathbf{#1}}

\def \pc {\,\mathrm{pc}} \def \kpc {\,\mathrm{kpc}}  
\def \kmpersec {\,\mathrm{km}\,\mathrm{s}^{-1}}

\def \degrees {^{\circ}}

\newcommand{\pdiff}[1]{\frac{\partial}{\partial #1}}

\def \half {\frac{1}{2}}

\newcommand{\text}[1] {\; \textrm{#1} \;}

\newcommand{\eqref}[1]{(\ref{#1})}

\newenvironment{remark*}{\textit{Remark.}}{}

\def\spose#1{\hbox to 0pt{#1\hss}}
\def\simlt{\mathrel{\spose{\lower 3pt\hbox{$\mathchar"218$}}
     \raise 2.0pt\hbox{$\mathchar"13C$}}}
\def\simgt{\mathrel{\spose{\lower 3pt\hbox{$\mathchar"218$}}
     \raise 2.0pt\hbox{$\mathchar"13E$}}}

\def \detA {\det \mathbf{A}}

\def \psixx {\psi_{xx}}  \def \psixy {\psi_{xy}}  \def \psiyy {\psi_{yy}}

 \def \vecxi {\bxi} 
\def \vecx {\vect{x}}   \def \vecr {\vect{r}}

\def \zl {z_{\mathrm{l}}}  \def \zs {z_{\mathrm{s}}}

\def \rc {r_{c}}  \def \rcOne {r_{c_{1}}}  \def \rcTwo {r_{c_{2}}}

\title[Lensing by Binary Galaxies]
{Lensing by Binary Galaxies Modelled as Isothermal Spheres}

\author[E.M. Shin and N.W. Evans] {E.M. Shin\thanks{E-mail:
ems@ast.cam.ac.uk; nwe@ast.cam.ac.uk} and
N.W. Evans\footnotemark[1]\\ Institute of Astronomy, University of
Cambridge, Madingley Road, Cambridge, CB3 0HA, United Kingdom}

\begin{document}

\pagerange{\pageref{firstpage}--\pageref{lastpage}} \pubyear{2008}

\maketitle
\label{firstpage}

\begin{abstract}
  We consider the problem of lensing by binary galaxies idealized as
  two isothermal spheres. This is a natural extension of the problem
  of lensing by binary point masses first studied by \citet{Sc86}.  In
  a wide binary, each galaxy possesses individual tangential, nearly
  astroidal, caustics and roundish radial caustics.  As the separation
  of the binary is made smaller, the caustics undergo a sequence of
  metamorphoses. The first metamorphosis occurs when the tangential
  caustics merge to form a single six-cusped caustic, lying interior
  to the radial caustics. At still smaller separations, the six-cusped
  caustic undergoes the second metamorphosis and splits into a
  four-cusped caustic and two three-cusped caustics, which shrink to
  zero size (an elliptic umbilic catastrophe) before they enlarge
  again and move away from the origin perpendicular to the binary
  axis. Finally, a third metamorphosis occurs as the three-cusp
  caustics join the radial caustics, leaving an inner distorted
  astroid caustic enclosed by two outer caustics. The maximum number
  of images possible is 7. Classifying the multiple imaging according
  to critical isochrones, there are only 8 possibilities: 2
  three-image cases, 3 five-image cases, and 3 seven-image cases. When
  the isothermal spheres are singular, the core images vanish into the
  central singularity. The number of images may then be 1, 2, 3, 4 or
  5, depending on the source location, and the separation and masses
  of the pair of lensing galaxies. The locations of metamorphoses, and
  the onset of threefold and fivefold multiple imaging, can be worked
  out analytically in this case.
\end{abstract}


\section{Introduction}

The Cambridge Sloan Survey of Wide Arcs in the Sky (CASSOWARY, see
Belokurov et al. 2007, 2008) has uncovered a number of examples of
gravitational lensing by luminous red galaxies (LRGs). For example,
the lens CASSOWARY 2 has two very bright and massive LRGs at redshifts
of $z = 0.426$ and $0.432$ respectively acting as lenses for a blue
star-forming galaxy at $z=0.970$. Although the LRGs lie in a loose
group, a natural starting point is that the lens is a close pair of
galaxies. CASSOWARY 5 has two early-type galaxies separated by $\sim
5^{\prime\prime}$ on the sky acting as lenses. The three images of the
high redshift source visible in the Sloan Digital Sky Survey data
straddle the two lenses.  CASSOWARY's search strategy has proved adept
at finding gravitational lenses with large separation images ($\gtrsim
3^{\prime\prime}$), which naturally enough are typically caused by two
or lensing galaxies.  There are also a number of existing strong
lenses for which there appears to be more than one lensing galaxy --
such as HE 0230-213~\citep{Wi99}, PMN J0134-0931~\citep{Wi03} and
B1359+154~\citep{Ru01}. In fact, in roughly a quarter of all strong
lenses, the lensing potential is known to be more complex than a
single lensing galaxy, with usually a pair or group of galaxies being
implicated.

The only binary lens that has so far received detailed scrutiny is the
important case of two point masses. \citet{Sc86} carried out a
detailed study of the caustics, critical curves and imaging
properties. In particular, they showed that the simpler model of a
point mass plus external shear \citep{Ch84,An06} is not always a
reliable description of the lensing properties of binary point
masses. Given the importance of microlensing surveys for planets,
there has been much subsequent theoretical work on the binary point
mass case~\citep[see e.g.,][]{Wi90,Ma91,Wi95,As02}. The theory has
borne fruit in detailed modelling of binary lens events, including the
first microlens mass determinations~\citep{An02}, and the discovery of
the first Neptune mass extra-solar planets~\citep{Be06}.

Here, we provide a theoretical treatment of the problem of lensing by
two isothermal spheres. The model has been considered numerically in
\citet{Ko88}, though primarily in the context of lensing in two planes
at different redshifts.  The motivation of our study is
threefold. First, a binary galaxy model has a direct application to
the close galaxy pair lenses found by the CASSOWARY survey or to
groups and clusters acting as lenses. Second, the investigation of the
properties of highly non-symmetric gravitational lenses is seriously
incomplete, and so the model provides an interesting counterweight to
the binary point mass case. Third, in the case of binary singular
isothermal spheres, many of the properties can be found analytically,
giving insights that are not so easy to obtain using numerical
work. The number of lenses for which analytic progress can be made is
still very small.

The paper is organised as follows.  In
Section~\ref{sec:BE-BCP-Curves}, we numerically examine the different
possible critical curve topologies of two cored isothermal
spheres. Different image configurations are explored by considering
the Fermat surfaces in Section~\ref{sec:BE-ImageGeometries}, and
Section~\ref{sec:BE-BSIS} deals with the special case of singular
isothermal spheres, and derives a number of exact results.

\section{Two Cored Isothermal Spheres}
\label{sec:BE-BCP-Curves}

\subsection{Models}

We use dimensionless source plane $\vecxi = (\xi, \eta)$ and lens
plane coordinates $\vecx = (x,y)$ ~\citep[see e.g.,][]{Sc92}. The two
isothermal spheres are centred at $\pm a$ on the $x$-axis. The
deflection potential is
\begin{eqnarray}\label{eq:BE-BCP-psi}
\psi(x,y) &=& E_{1} \: \left[ \rcOne^{2} + (x + a)^{2} + y^{2}
\right]^{1/2} \nonumber \\
&+& E_{2} \: \left[ \rcTwo^{2} + (x - a)^{2} + y^{2} \right]^{1/2},
\end{eqnarray}
where $\rcOne$ and $\rcTwo$ are the core radii of the isothermal
spheres. If the core radii vanish, this reduces to the potential of
two singular isothermal spheres with Einstein radii $E_{1}$ and
$E_{2}$. We take $E_{1} \leq E_{2}$ and $a > 0$ without loss of
generality.

Let us introduce $r_{1}^2 = (a+x)^2 + y^2$ and $r_{2}^2 = (a-x)^2 + y^2$. Then
the convergence is
\begin{equation}\label{eq:BE-Models-kappa}
\kappa(\vecx) = \frac{E_1}{2} \frac{ 2\rcOne^2 + r_{1}^2}{(\rcOne^2 +
  r_{1}^2)^{3/2}}+ \frac{E_2}{2}\frac{2\rcTwo^2 + r_{2}^2}{(\rcTwo^2 + r_{2}^2)^{3/2}}.
\end{equation}
whilst the shear components are
\begin{subequations}\label{eq:BE-Models-gamma-components}
\begin{eqnarray}
\gamma_1(\vecx) &=& \frac{E_1}{2}\frac{ y^2 - (a+x)^2}{(\rcOne^2 + r_{1}^2)^{3/2}}
+\frac{E_2}{2}\frac{ y^2 - (a-x)^2}{(\rcTwo^2 + r_{2}^2)^{3/2}},\\
\gamma_2(\vecx) &=& E_2\frac{(a-x) y}{(\rcTwo^2 + r_{2}^2)^{3/2}} - E_1\frac{(a+x) y}{(\rcOne^2 + r_{1}^2)^{3/2}}.
\end{eqnarray}
\end{subequations}
The Jacobian of the lens mapping may be found either from
\begin{equation} \label{eq:BE-Models-detA-Cartesian-psis}
\detA(\vecx) = \left[1 - \psixx(\vecx)\right]\left[1 - \psiyy(\vecx)\right] - \psixy(\vecx)^{2} \,,
\end{equation}
or from
\begin{equation} \label{eq:BE-Models-detA-kappa-gamma}
\detA(\vecx) = \left[1 - \kappa(\vecx)\right]^{2} - \gamma(\vecx)^{2}\,,
\end{equation}
where the magnitude of shear is
\begin{equation} \label{eq:BE-Models-ShearMagnitude-FromComponents}
\gamma(\vecx) = \sqrt{\gamma_{1}(\vecx)^{2} + \gamma_{2}(\vecx)^{2}} \:.
\end{equation}
For an isolated, singular isothermal sphere, $\gamma = \kappa$ and so
we obtain the familiar result $\detA = 1 - 2\kappa$. The analogue of
this in the binary, singular case is
\begin{equation}\label{eq:BE-Models-Wyn}
\detA = 1 - 2\kappa +4 E_1E_2 \frac{a^2y^2}{r_{1}^3r_{2}^3}
\end{equation}
and so $\detA = 1 - 2\kappa$ on the axis of two singular isothermal
spheres.

\subsection{Two Identical Galaxies with Varying Separation}\label{secsub:BE-CCs-TwoIdenticalSpheres}

\begin{figure*}
\vspace*{-2.0cm}
\epsfxsize = 12.5cm \centreline{\epsfbox{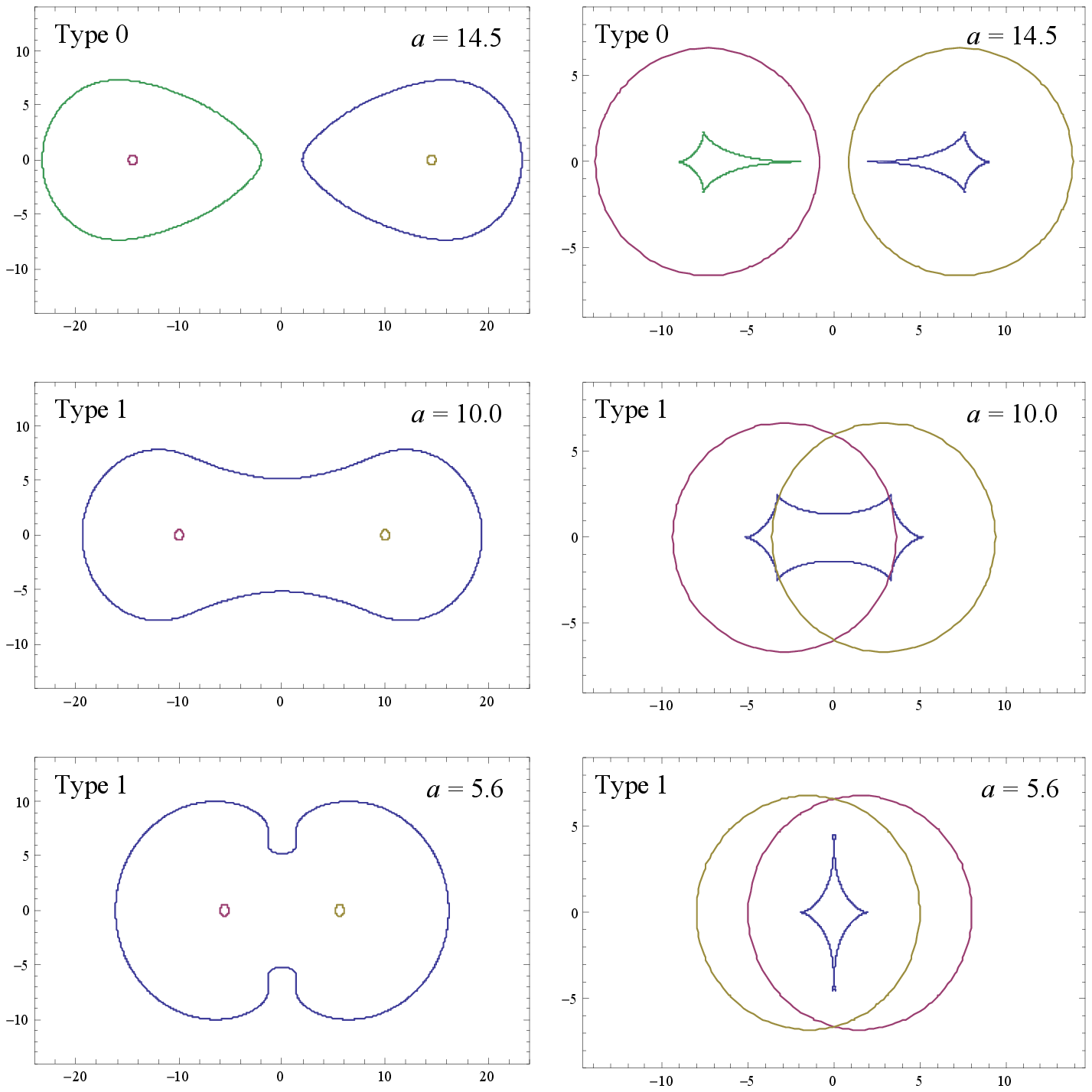}}
\vspace{0.2cm}
\epsfxsize = 12.5cm \centreline{\epsfbox{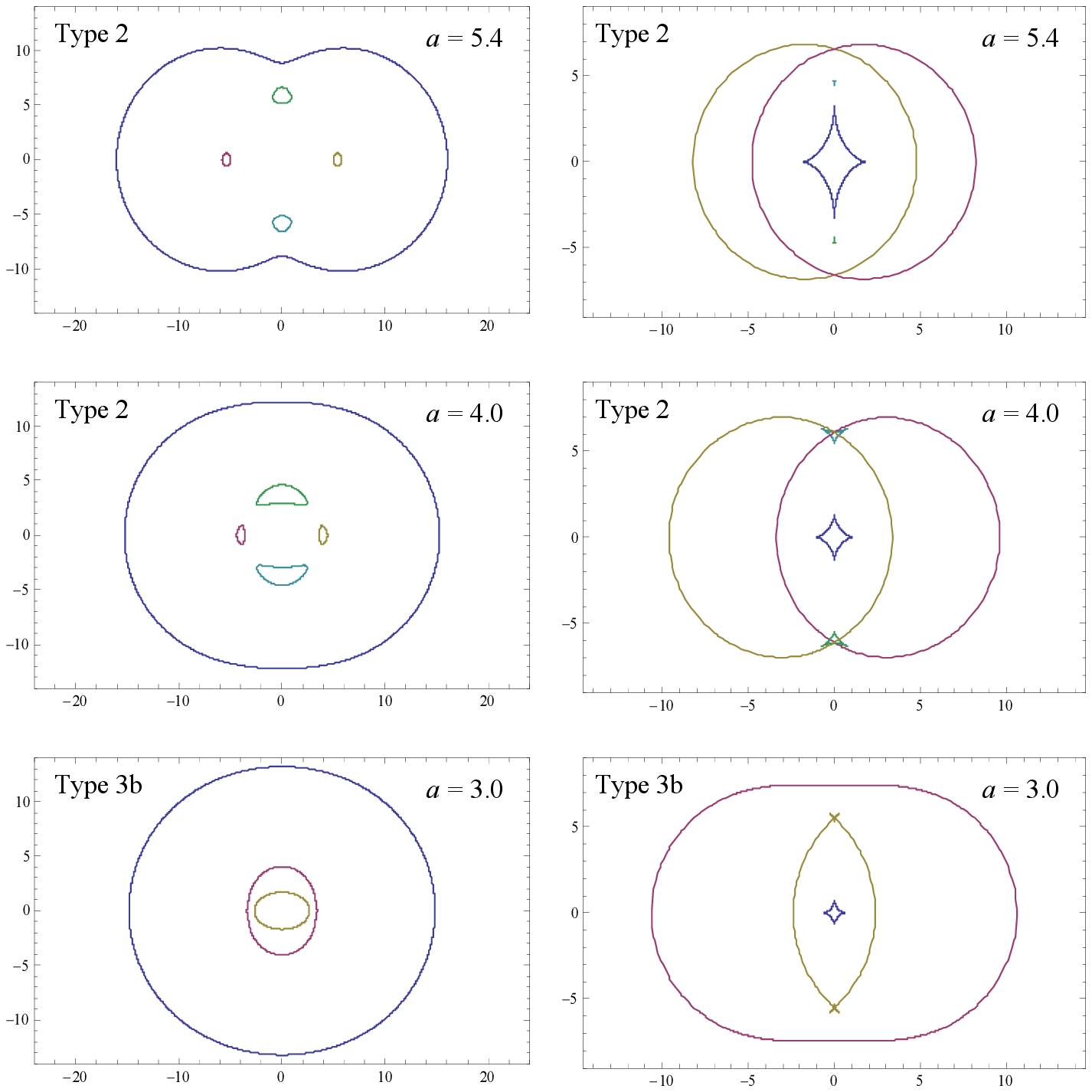}}
\caption{Critical curves (left panels) and caustics (right panels) for
  two identical fiducial isothermal spheres (with dimensionless
  Einstein radius $\sim\!7.11$) separated by dimensionless distance
  $2a$. Colours of corresponding critical curves and caustics
  match. The \textit{Type} of the critical-curves or caustics is
  defined in the text. \label{fig:BE-BCP-Set-A}}
\end{figure*}

Let us take as properties of a fiducial elliptical galaxy lens a
velocity dispersion $\sigma = 250\kmpersec$, a core radius
$\rcOne=\rcTwo = 100\pc$, and a redshift $\zl = 0.46$. Let us take the
source redshift as $\zs = 2.15$, and use a $\Lambda CDM$ concordance
cosmology. As shown in \citet{Sh08}, this choice of parameters is
astrophysically reasonable. The corresponding dimensionless Einstein
radius is $E \approx 7.11$ (where the length scale is chosen to be
$1\kpc$). We place two such fiducial isothermal spheres at $(-a,0)$
and $(+a,0)$, and numerically find critical curves and caustics for
varying $a$. The different possible configurations are shown in
Fig.~\ref{fig:BE-BCP-Set-A}.

When $a$ is large (top-most panels of Fig.~\ref{fig:BE-BCP-Set-A}),
the critical curves and caustics of the two lenses are disjoint, but
the outer tangential critical curves are not circular (as they would
be for a single such lens in isolation) and the corresponding caustics
are distorted astroids with four cusps. There are small radial
critical curves around each lens that map to large roundish
caustics. We call this configuration of critical curves, and the
corresponding caustic configuration, `Type 0'. For $a = 10.0$, the
tangential critical curves have merged into a single outer critical
curve (which we call the common tangential critical curve), and the
corresponding caustics have merged to form a single six-cusp
caustic. We call this arrangement `Type 1'. The six-cusp caustic
elongates along the $y$-axis as $a$ decreases further, and two small
three-cusp caustics split off, as two small critical curves
(henceforth `bean' critical curves) pinch off from the outermost
critical curve, to leave a single four-cusp astroid caustic ($a = 5.6$
and $ a = 5.4$). We call this `Type 2'. The physical origin of the
`bean' critical curves is explained later
(\S~\ref{secsub:BE-BSIS-EUcatastrophe}) in the context of singular
isothermal spheres. At $a = 4.0$, the three-cusp caustics have moved
away from the origin and enlarged, whilst the `bean' critical curves
have moved towards the origin and enlarged. Decreasing $a$ still
further, the two `bean' critical curves merge with the two radial
critical curves, creating two critical curves around the origin
(bottom-most panels). The inner one maps to a caustic with two
butterfly cusps, whilst the outer one maps to the outermost
caustic. The outermost critical curve still corresponds to an astroid
caustic around the origin. We call this configuration `Type 3b'.

\subsection{Two Galaxies with Different Einstein Radii $E_{1}$
  and $E_{2}$ }\label{secsub:BE-BCP-CCs-E1-E2-Space}

Although there are five free parameters in \eqref{eq:BE-BCP-psi}, the
geometry of the critical curves and caustics are determined by
four. This is most easily seen from the lens equation
\begin{subequations}\label{eq:BE-LensEquation}
\begin{equation}
\xi = x \;-\; E_{1}\,\frac{(x+a)}{\left(\rcOne^{2} + r_{1}^{2}\right)^{1/2}} \;-\; E_{2}\,\frac{(x-a)}{\left( \rcTwo^{2} + r_{2}^{2}\right)^{1/2}}
\end{equation}
\begin{equation}
\eta = y \left[ 1\;-\; \frac{E_{1}}{\left(\rcOne^{2} + r_{1}^{2}\right)^{1/2}} \:-\: \frac{E_{2}}{\left(\rcTwo^{2} + r_{2}^{2}\right)^{1/2}} \right].
\end{equation}
\end{subequations}
If all five parameters are scaled by $\lambda$, the new lens has the
same critical curve and caustic geometry, just on a different scale:
choosing new coordinates $(x\dash,y\dash) = \lambda^{-1}(x,y)$,
$(\xi\dash,\eta\dash) = \lambda^{-1}(\xi,\eta)$ reduces the new lens
equation to the old one.

Now, the most physically interesting regime has $\rc \ll a$ -- that
is, the separation between the two lensing galaxies is much larger
than their core radii -- whilst $E_{1}, E_{2}$ and $a$ are
comparable. The scaling degeneracy means that it is only the ratios
$E_{1}/a$ and $E_{2}/a$ that determine the geometry of the critical
curves and caustics (up to the effect of the core radii). So we fix
$a$ (to 10, specifically), allow $E_{1}$ and $E_{2}$ to vary, and find
which critical curve geometries arise from different parts of the
$E_{1}$-$E_{2}$ parameter plane. Since only small core radii are
physical, we do not explore the $\rcOne$ and $\rcTwo$ dimensions of
parameter space much. For simplicity, $r_{c1,c2}$ are set to scale
with $E_{1,2}$ as $(0.1/7.11)E_{1,2}$ respectively, so that a doubling
of both $E_{i}$ corresponds, physically, to halving the separation
$2a$ between the lens galaxies. (The factor $(0.1/7.11)$ normalizes
the core radii to the fiducial isothermal spheres.)

\begin{figure*}
\epsfxsize = 14.5cm \centreline{\epsfbox{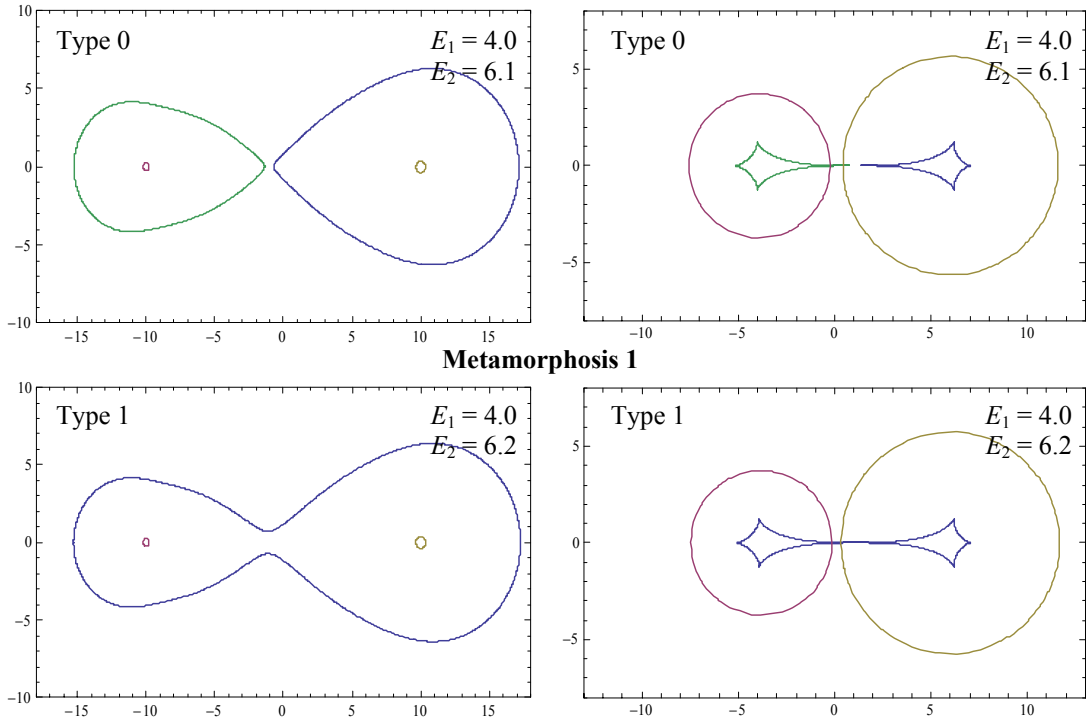}}
\caption{Critical curves (left panels) and caustics (right panels) of
  Metamorphosis 1, which takes us from Type 0 (upper panels) to Type 1
  (lower panels) morphologies. $E_{1}$ and $E_{2}$ in the lower panels
  are greater than or equal to $E_{1}$ and $E_{2}$ in the upper
  panels. As in the case of identical isothermal spheres
  (Fig.~\ref{fig:BE-BCP-Set-A}), the two tangential critical curves
  merge, as do the corresponding distorted four-cusp astroid
  caustics. \label{fig:BE-BCP-Metamorphosis-1}}
\end{figure*}

\begin{figure*}
\epsfxsize = 14.5cm \centreline{\epsfbox{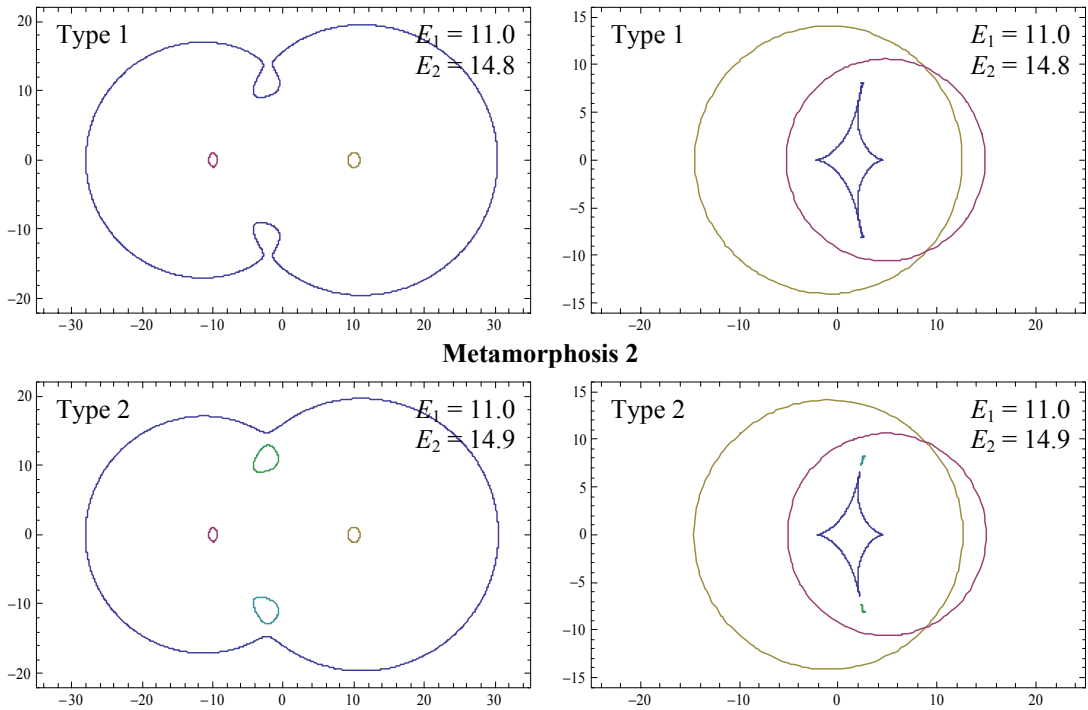}}
\caption{Critical curves (left panels) and caustics (right panels) of
  Metamorphosis 2, which takes us from Type 1 (upper panels) to Type 2
  (lower panels) morphologies. $E_{1}$ and $E_{2}$ in the lower panels
  are greater than or equal to $E_{1}$ and $E_{2}$ in the upper
  panels.  Two critical curves pinch off from the outer critical
  curve, corresponding to two three-cusp caustics pinching off from
  the hexacuspid caustic. \label{fig:BE-BCP-Metamorphosis-2}}
\end{figure*}

\begin{figure*}
\epsfxsize = 14.5cm \centreline{\epsfbox{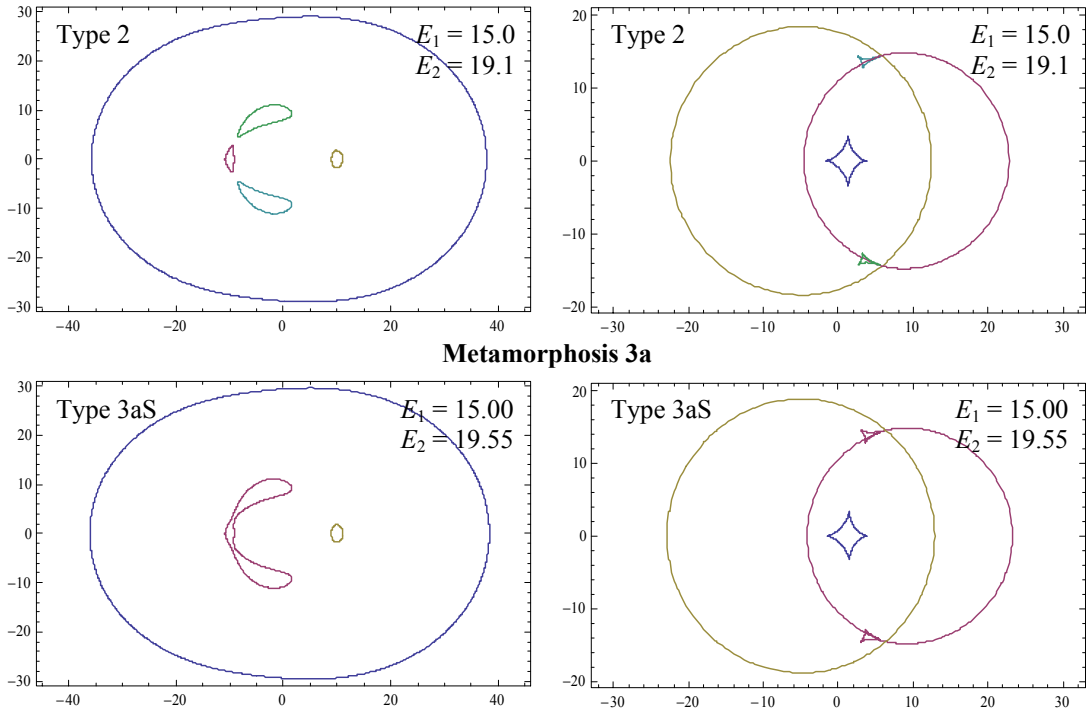}}
\caption{Critical curves (left panels) and caustics (right panels) of
  Metamorphosis `3a'. $E_{1}$ and $E_{2}$ in the lower panels are
  greater than or equal to $E_{1}$ and $E_{2}$ in the upper
  panels. (Metamorphoses 3a and 3b are distinct if $E_{1} \neq
  E_{2}$.) In 3a, the two critical curves corresponding to the
  three-cusp caustics merge with one of the two small `radial'
  critical curves to form a `Pacman' critical curve. The three-cusp
  caustics merge with the corresponding radial caustic, giving it six
  cusps, although increasing $E_{2}$ further can reduce that to two
  cusps via a swallowtail folding (see
  Fig.~\ref{fig:BE-BCP-Swallowtail-Met}). \label{fig:BE-BCP-Metamorphosis-3a}}
\end{figure*}

\begin{figure*}
\epsfxsize = 14.5cm \centreline{\epsfbox{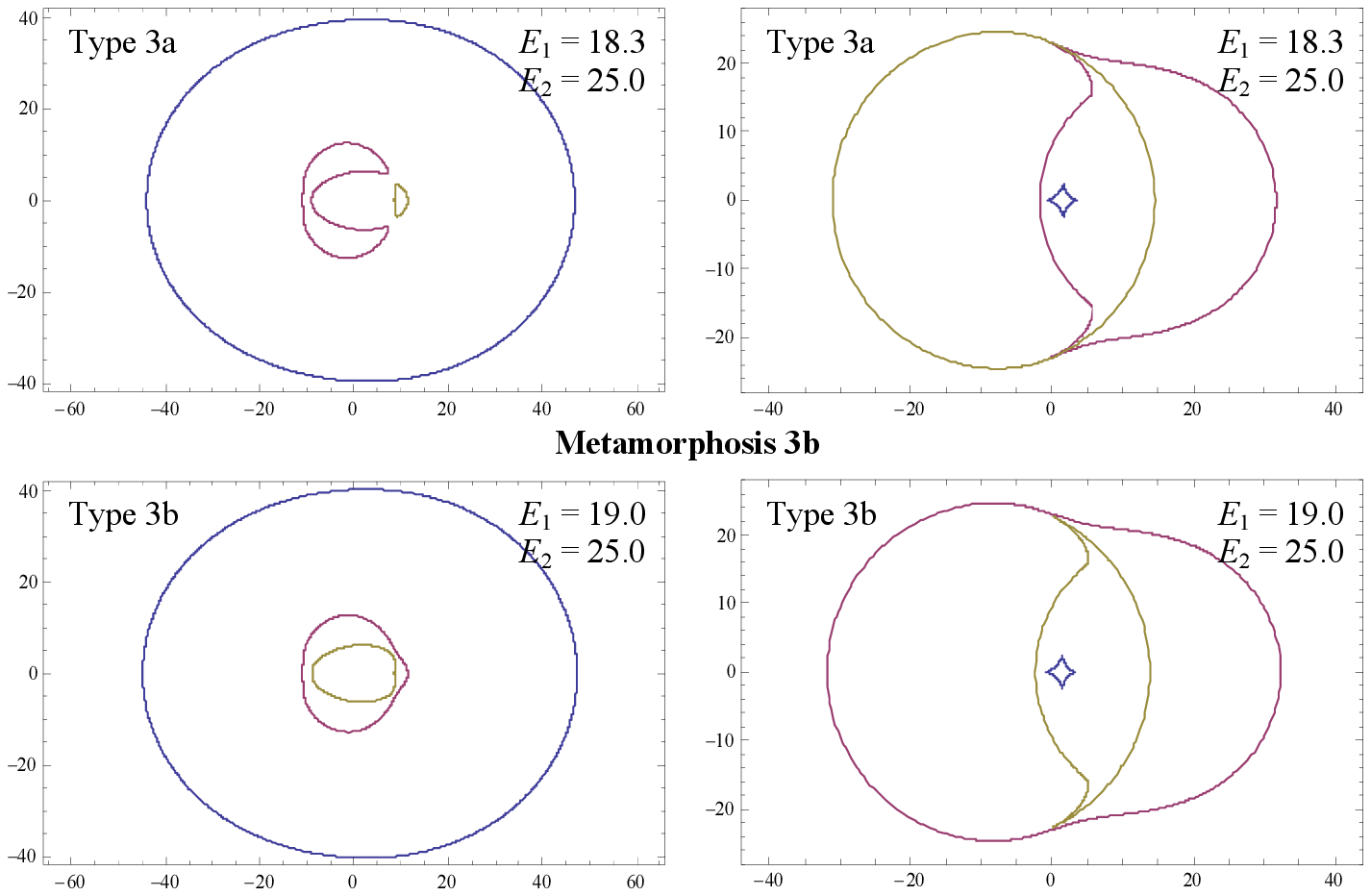}}
\caption{Critical curves (left panels) and caustics (right panels) of
  Metamorphosis `3b'. $E_{1}$ and $E_{2}$ in the lower panels are
  greater than or equal to $E_{1}$ and $E_{2}$ in the upper
  panels. The `Pacman' critical curve merges with the remaining radial
  critical curve; the caustics change accordingly. (Swallowtails --
  see Fig.~\ref{fig:BE-BCP-Swallowtail-Met} -- are not visible with
  the curves plotted on this
  scale.) \label{fig:BE-BCP-Metamorphosis-3b}}
\end{figure*}

We find that there are five possible critical curve topologies. There
are the four seen in \S~\ref{secsub:BE-CCs-TwoIdenticalSpheres}: Type
0 (disjoint tangential critical curves), Type 1 (a common tangential
critical curve), Type 2 (a common tangential critical curve
surrounding two `bean' critical curves in addition to the two radial
critical curves), and Type 3b (the common tangential critical curve,
and two other critical curves in a distorted annulus around the
origin, with the galaxy centres lying within the edges of the
annulus). In addition, there is a Type 3a topology: unequal isothermal
spheres mean that the two `bean' critical curves merge with one of the
radial critical curves before the other. These five critical curve
topologies are separated by four metamorphoses of critical curve
geometry, as shown in Figures~\ref{fig:BE-BCP-Metamorphosis-1}
to~\ref{fig:BE-BCP-Metamorphosis-3b}, which we call Metamorphoses 1,
2, 3a and 3b. Metamorphosis 1 marks the transition from Type 0 to Type
1 critical curves, and so on. Recall when viewing these figures that $E_{i}$ are always dimensionless Einstein radii (in the same units as the axes) and that the isothermal spheres are centred at $\pm 10$ on the $x$-axis.

\begin{figure*}
\epsfxsize = 14.5cm \centreline{\epsfbox{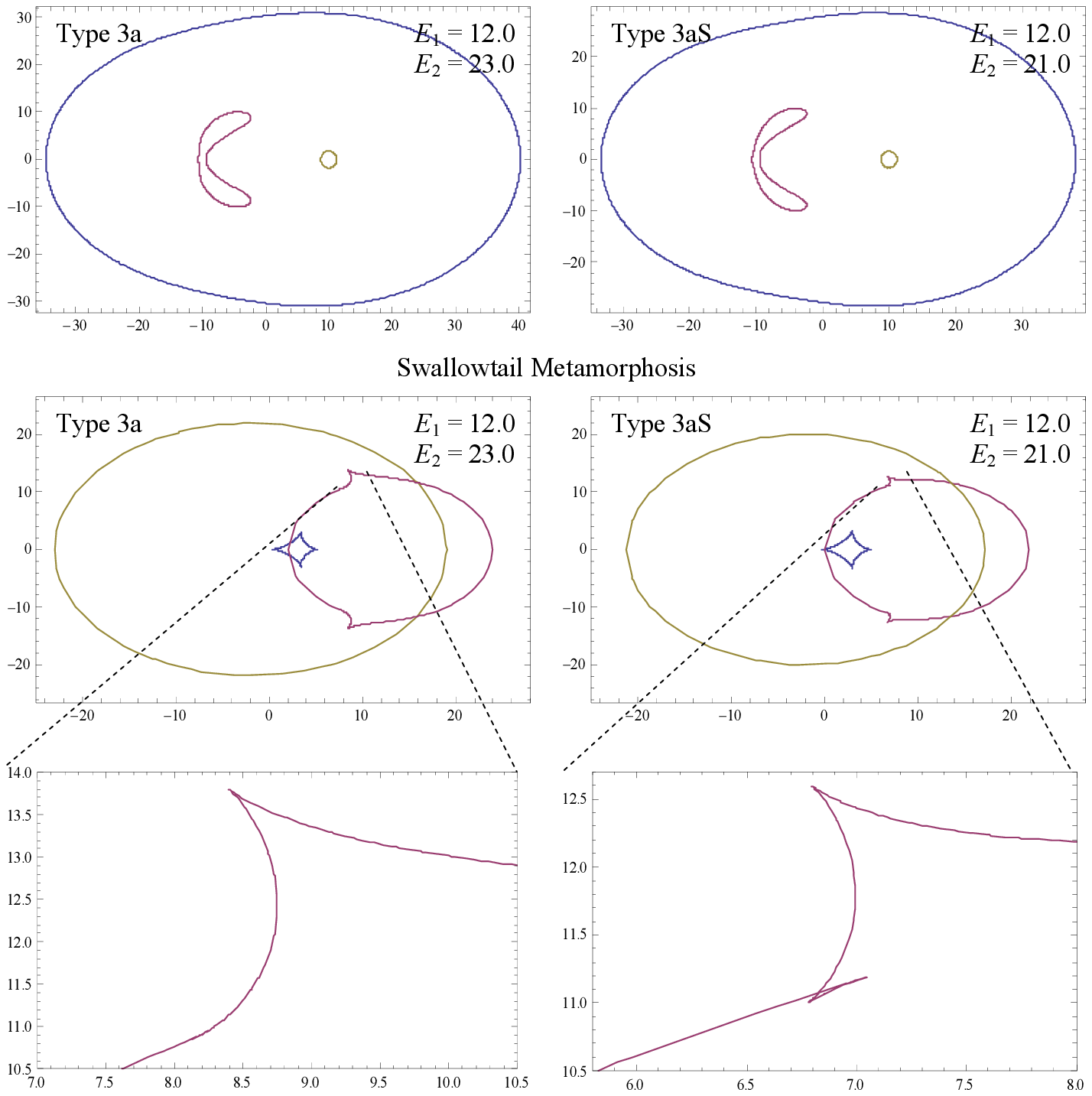}}
\caption{The Swallowtail Metamorphosis. Two of the three cusps in this
  caustic belong to a swallowtail unfolding. Note that critical curves
  are shown in the two \textit{top} panels and caustics in the lower
  ones. $E_{1}$ and $E_{2}$ in the right panels are less than or equal
  to $E_{1}$ and $E_{2}$ in the left
  panels. \label{fig:BE-BCP-Swallowtail-Met}}
\end{figure*}

The metamorphoses of caustics occur at metamorphoses of the critical
curves, except for the development of swallowtails in one of the
caustics without any corresponding change in the topology of the
critical curves: see Fig.~\ref{fig:BE-BCP-Swallowtail-Met}. (For the
mathematical background to catastrophe theory, see e.g., \citealp{Ar86},
whilst for an introduction to critical-curve and caustic
metamorphoses, see e.g. \citealp{PLW}.) So we also call caustics Type
0, 1 and so on if the critical curves are of that type. Note, however,
an astroid caustic can still pierce a radial caustic, for example,
without the critical curve topology changing. But caustics do not
merge or split unless the critical curves do.


In Type 0 critical curves, $\detA > 0$ outside the tangential critical
curves, and $\detA < 0$ within them except for the small regions
within the radial critical curves around the galaxy centres at $(\pm
a, 0)$. There is a saddle of $\detA$ on the $x$-axis, and as $E_{i}$
are increased, $\detA$ at the saddle decreases. As the saddle
decreases through zero, the two regions of negative $\detA$ link up,
forming Type 1 critical curves (Metamorphosis 1,
Fig~\ref{fig:BE-BCP-Metamorphosis-1}). The corresponding metamorphosis
in the caustics is the merging of the two four-cusp astroids, via a
`beak-to-beak' transition~\citep[see e.g.,][ Chap. 6]{Sc92}, into a
six-cusped curve, which we henceforth call a `hexacuspid'. Note also
that the hexacuspid can pierce the radial caustics.

As $E_{i}$ are increased further, the outer, common tangential
critical curve loses its bottleneck, but eventually develops dimples
around two small regions that have more positive $\detA$ than their
surroundings (these are regions where the shear is small: see
\S~\ref{secsub:BE-BSIS-EUcatastrophe}). Meanwhile, the hexacuspid
becomes less elongated along the $x$-axis and more elongated in the
$y$-axis. Increase $E_{i}$ still more and Metamorphosis 2 occurs
(Fig.~\ref{fig:BE-BCP-Metamorphosis-2}): the dimple in the
tangential critical curve closes completely around the small regions
of positive $\detA$, creating two `bean' critical curves. The two
small three-cusp caustics split off from the tangential caustic via
beak-to-beak metamorphoses.

At still larger $E_{i}$, the `bean' critical curves extend towards the
radial critical curves whilst the three-cusp caustics move outwards
towards the radial caustics. When $E_{2} \lesssim 20$, we have
Metamorphosis 3a (Fig.~\ref{fig:BE-BCP-Metamorphosis-3a}: the beans
merge with the left-hand radial critical curve, forming what we call
the `Pacman' critical curve~\footnote{Named after the characteristic
  shape of the protagonist in the eponymous computer game.}, and the three-cusp
caustics tack on to the corresponding radial caustic). As $E_{1}$
reaches $\lesssim 20$, the `Pacman' critical curve merges with the
remaining radial critical curve (Metamorphosis 3b,
Fig.~\ref{fig:BE-BCP-Metamorphosis-3b}) to form an annular region of
positive $\detA$, as the caustics change accordingly (via hyperbolic
umbilic metamorphoses again).

%

\begin{figure*}
\epsfxsize = 12cm \centreline{\epsfbox{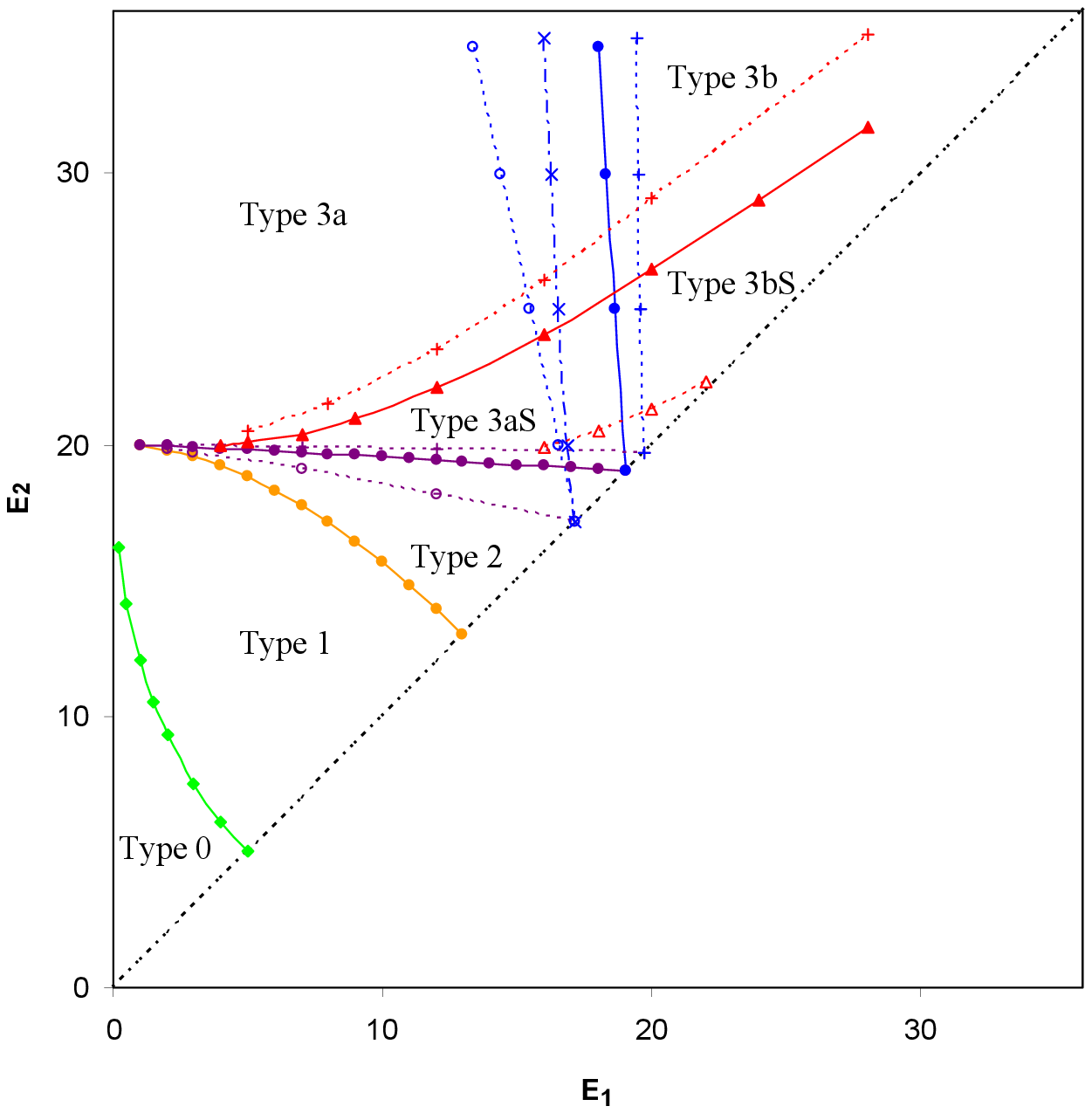}}
\caption{The critical curve metamorphoses divide $E_{1}$-$E_{2}$
  parameter space into regions of different critical curve and caustic
  geometries. Metamorphoses 1, 2, 3a and 3b are shown in green,
  orange, purple and blue, respectively. The swallowtail metamorphosis
  is shown in red. This plot is for $a = 10$, which is arbitrarily
  chosen: recall that the critical curves are determined (up to
  rescaling of axes) by the ratios $E_{1}/a$, $E_{2}/a$, $\rcOne/a$
  and $\rcTwo/a$. Solid points and lines correspond to the fiducial
  scaling of the core radius $r_{c\,1,2} = (0.1/7.11)E_{1,2}$ (see
  \S\ref{secsub:BE-BCP-CCs-E1-E2-Space}), whilst open points are for
  $r_{c\,1,2}$ scaling as $5\times(0.1/7.11)E_{1,2}$ and plus signs
  for $(1/5)\times(0.1/7.11)E_{1,2}$. The blue crosses are for
  Metamorphosis 3b (the metamorphosis most sensitive to core radii)
  with $r_{c\,1,2}$ fixed instead of scaling with
  $E_{i}$. Metamorphoses 1 and 2 are not significantly dependent on
  $r_{c\,1,2}$, whilst the other metamorphoses (especially 3b and the
  swallowtail) are very sensitive to them. Increasing $r_{c\,1,2}$
  suppresses formation of swallowtail cusps and brings metamorphoses
  3a and 3b to smaller $E_{1,2}$. The lines of metamorphoses 3a and 3b
  would approach $E_{2} = 2a$ ($= 20$ here) and $E_{1} = 2a (= 20)$ if
  $\rcOne \to 0$, $\rcTwo \to 0$ (see
  \S~\ref{secsub:BE-BSIS-Met3ab}). \label{fig:BE-BCP-E1-E2-plane}}
\end{figure*}

The five critical curve regimes (Types 0, 1, 2, 3a and 3b), into which
the $E_{1}$-$E_{2}$ parameter plane is divided by the four critical
curve metamorphoses, are shown in
Fig.~\ref{fig:BE-BCP-E1-E2-plane}. Recall that we have fixed $a = 10$
because the critical curve and caustic geometry is dependent mainly on
the ratios $E_{1,2}/a$ and $r_{c\,1,2}/a$ in the astrophysically
important regime of $a \! \gg \! r_{c\,1,2}$.  For small core radii, there is
a good analytic approximation (which is exact for singular isothermal
spheres -- see \S~\ref{secsub:BE-BSIS-Met1}) for the lens parameters
$E_{1}, E_{2}, a$ at which Metamorphosis 1 (the merging of the two
tangential critical curves) occurs. The other metamorphoses are
determined by numerically finding critical curves and caustics for
various $E_{i}$. Metamorphoses 3a and 3b depend very strongly on
$r_{c1,c2}$, whilst Metamorphoses 1 and 2 don't, as might be expected.

Higher order catastrophes such as swallowtails have been noted before
in a variety of lens models ~\citep{Ke00,Ev01,Br04}. Swallowtails can
form for Types 3a and 3b; regions in the $E_{1}$-$E_{2}$ plane where
the caustics have swallowtails are denoted by 3aS and 3bS. The
lower-right of Fig.~\ref{fig:BE-BCP-E1-E2-plane}, which is left blank,
is -- by the symmetry of the lens -- just the reflection of the
upper-left part of the plot in the line $E_{1} = E_{2}$ (shown
dotted).

From Fig.~\ref{fig:BE-BCP-E1-E2-plane}, we confirm that Type 0, 1, and
2 curves correspond to progressively stronger isothermal sphere
lenses. The size of the core radii affects the size of the radial
critical curves and hence when the radial and `bean' critical curves
merge via Metamorphoses 3a and 3b. As expected, increasing the core
radii mean that the mergers take place at smaller $E_{i}$, as the
radial critical curves enlarge (reaching out towards the bean critical
curves). The reason that smaller core radii enlarge the swallowtail
region of parameter space is that $r_{c} = 0$ singular isothermal
spheres always have three-cusp caustics (see \S~\ref{sec:BE-BSIS}).


It is also worth noting that the `bean' critical curves and the
corresponding three-cusp caustics pass through an elliptic umbilic
catastrophe in parameter space (where they shrink to zero size),
in-between Metamorphoses 2 and 3a. The condition on $E_{1}, E_{2}$ and
$a$ for this can be found exactly for the case $\rcOne = \rcTwo = 0$
(see \S~\ref{secsub:BE-BSIS-EUcatastrophe}), and remains a good
approximation for small $r_{c}$.

\section{Image geometries}\label{sec:BE-ImageGeometries}
\begin{figure*}
\epsfxsize = 10cm \centreline{\epsfbox{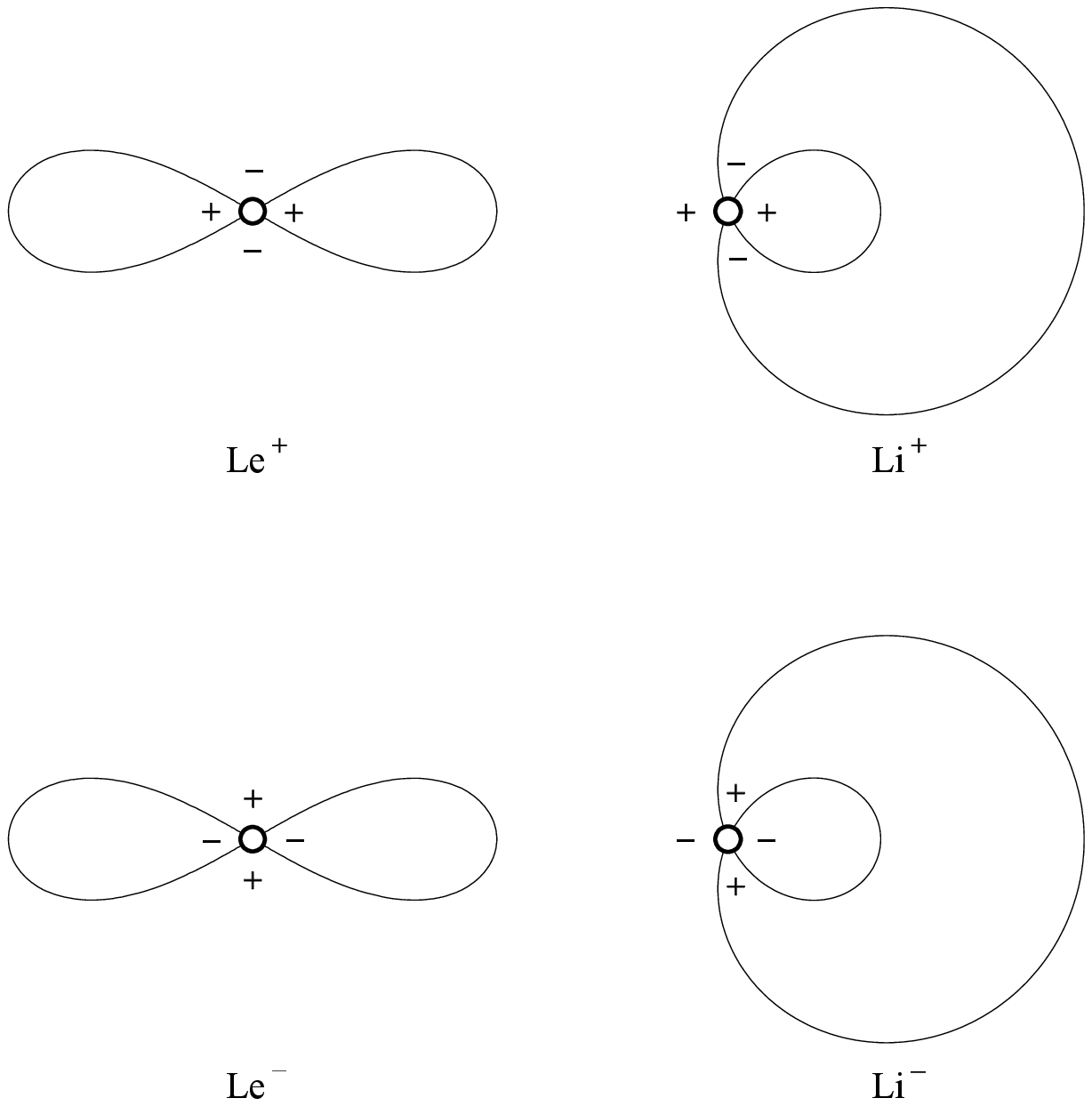}}
\caption{The four ways to close the critical isochrones of a single
  saddle point. The left panels show lemniscates, the right panels
  lima\c{c}ons. Critical isochrone topologies for multiple saddles are
  constructed by combining these basic shapes. \label{fig:BE-CritIso-LeLi-diagrams}}
\end{figure*}
\begin{figure*}
\epsfxsize = 12.5cm \centreline{\epsfbox{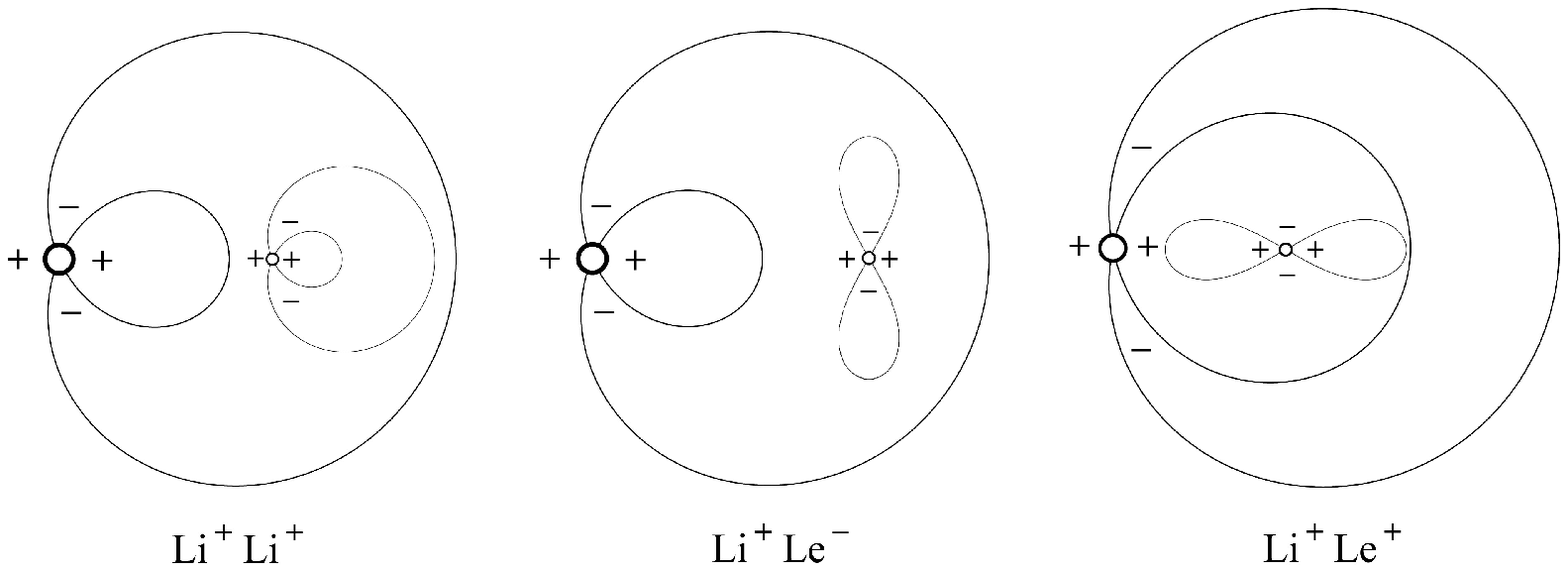}}
\vspace{0.5cm}
\epsfxsize = 12.5cm \centreline{\epsfbox{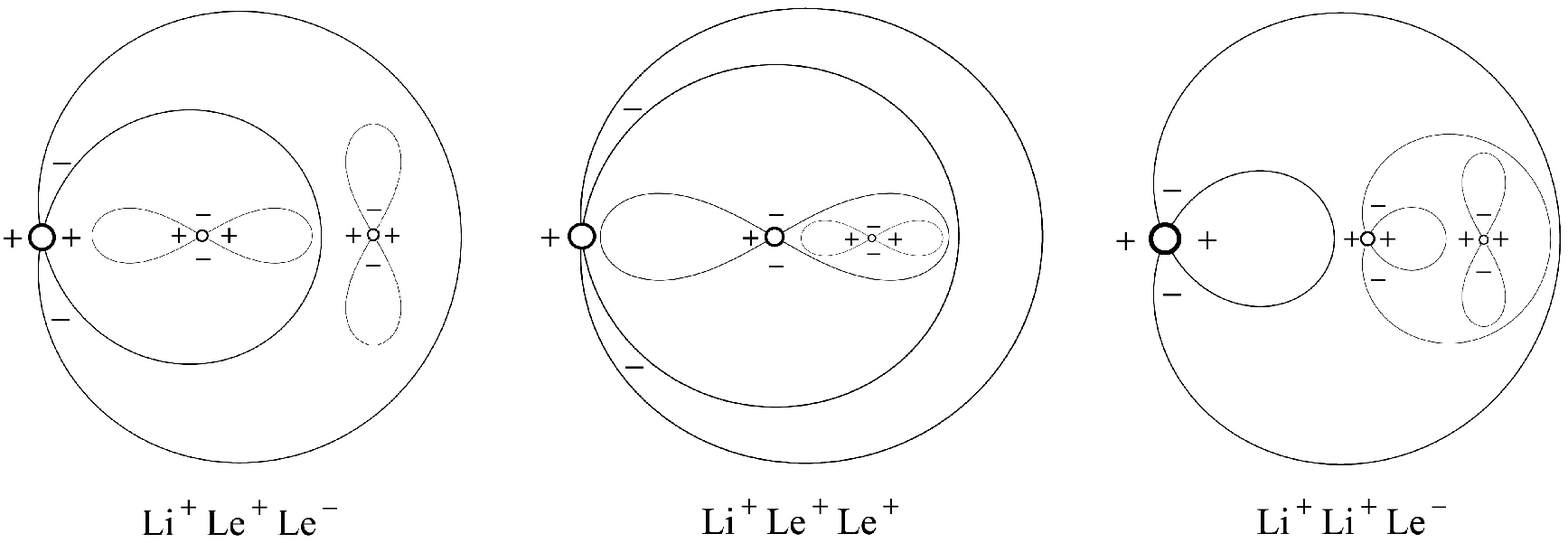}}
\caption{The five- and seven-image critical isochrone topologies that
  occur for the double isothermal sphere lens. They are named
  according to the shape of the critical isochrones (see
  Fig.~\ref{fig:BE-CritIso-LeLi-diagrams}), starting from the
  outermost. \label{fig:BE-BCP-FSTopologyDiagrams}}
\end{figure*}

\begin{figure*}
\epsfxsize = 8.4cm
\flushright{\epsfbox{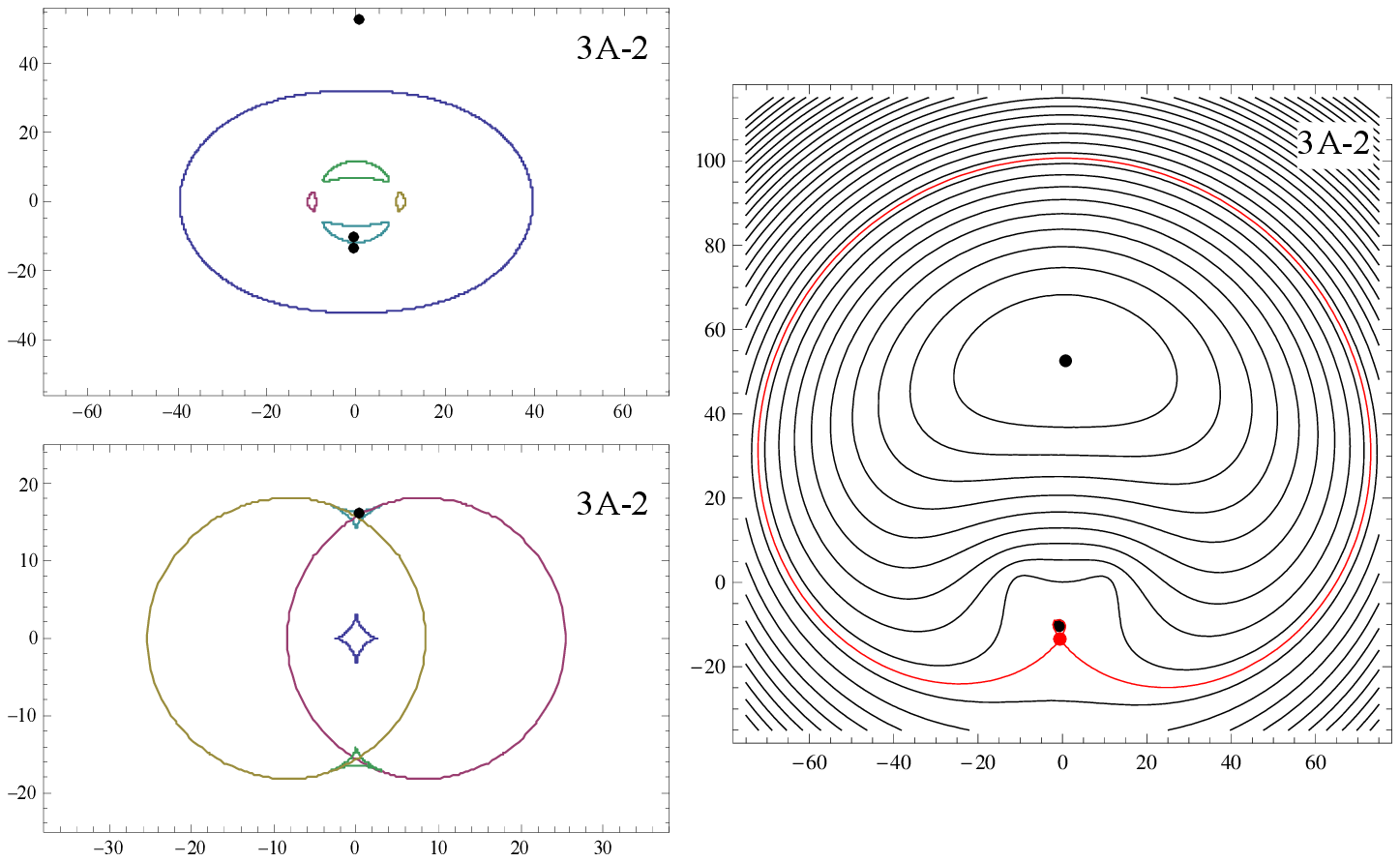}}
\vspace{-8cm}\flushleft{\epsfbox{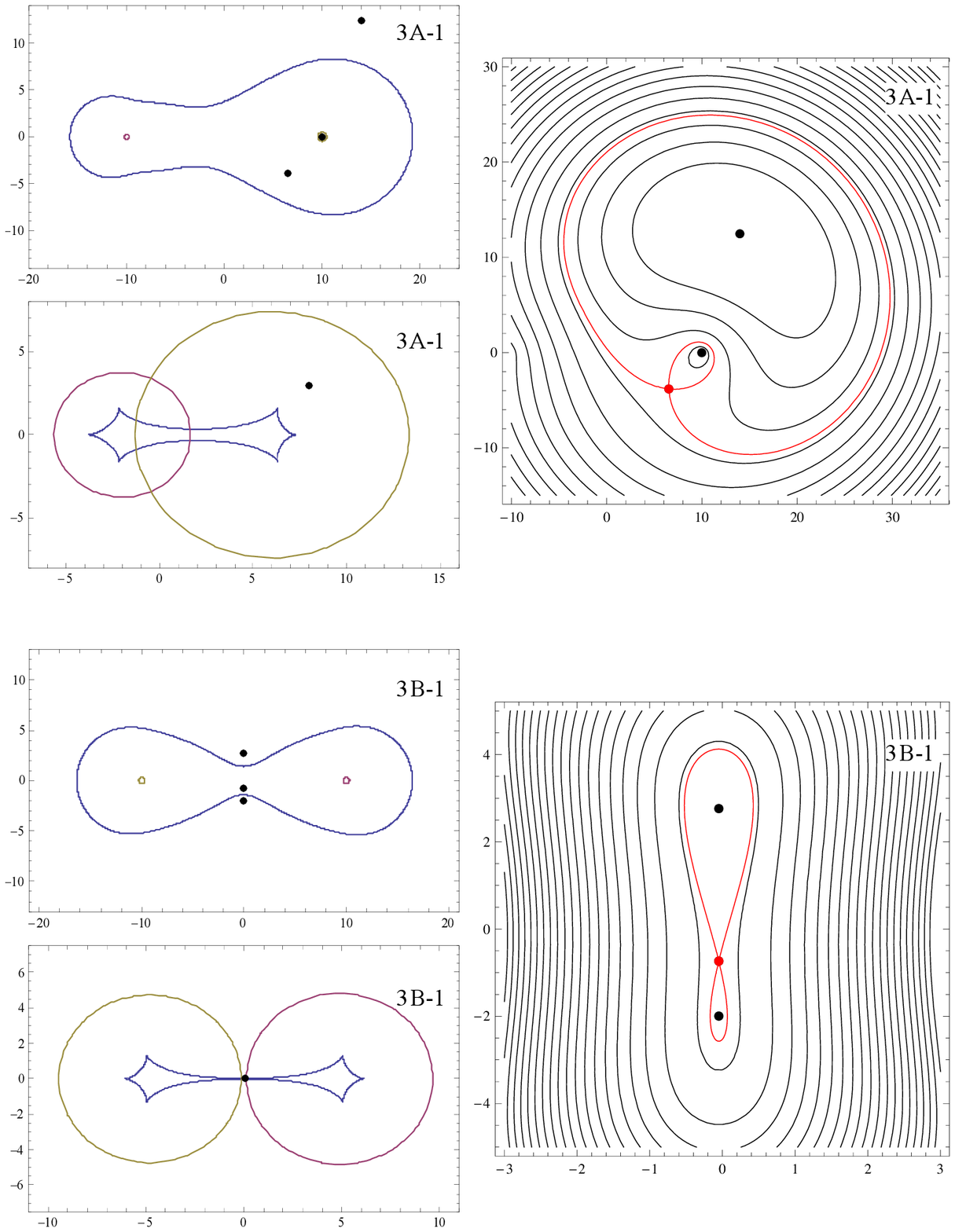}}
\caption{The different three-image Fermat surface topographies and
  image configurations. The left-hand panels show critical curves and
  image positions (upper panels of each example), and caustics and the
  position of the point source (lower panels). Colours of
  corresponding critical curves and caustics match. The right-hand
  panels show contours of the Fermat potential together with the
  saddle-points and their critical isochrones (red points and
  contours) as well as the images corresponding to the maximum and the
  minimum in the Fermat potential (black points). In example 3A-1, one
  of the three images is a highly-demagnified core image, whilst in
  3B-1 all three images are of high magnification. The source in 3A-2
  is inside a three-cusp caustic but outside both radial caustics. The
  critical isochrone is a lima\c{c}on, as in 3A-1, but this system
  would be observed as a triplet: the two images near the `bean'
  critical curve are of high magnification, and the primary image is
  also magnified. \label{fig:BE-BCP-FS_3ims}}
\end{figure*}

\begin{figure*}
\epsfxsize = 8.4cm 
\flushright{\epsfbox{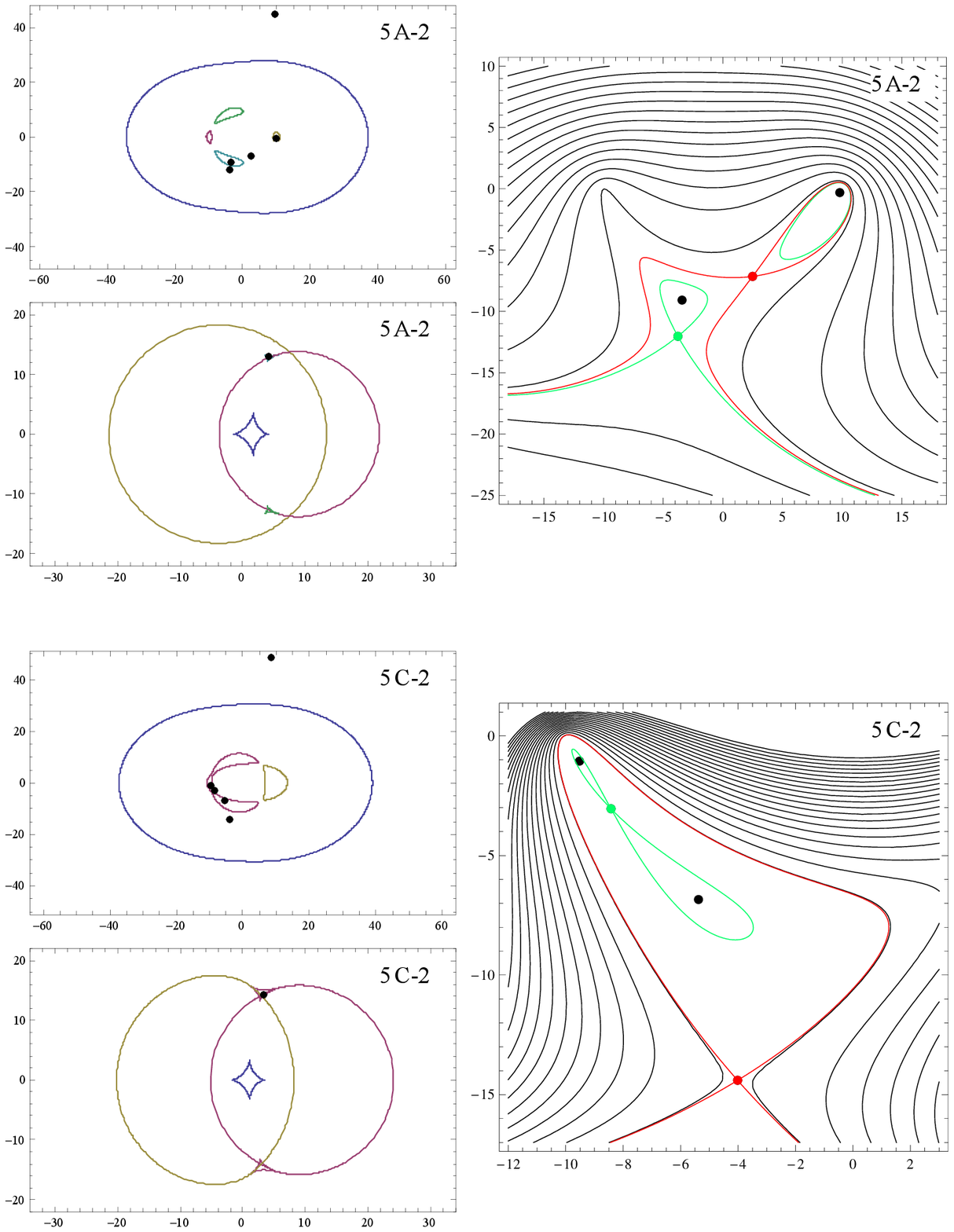}}
\vspace{-13cm}\flushleft{\epsfbox{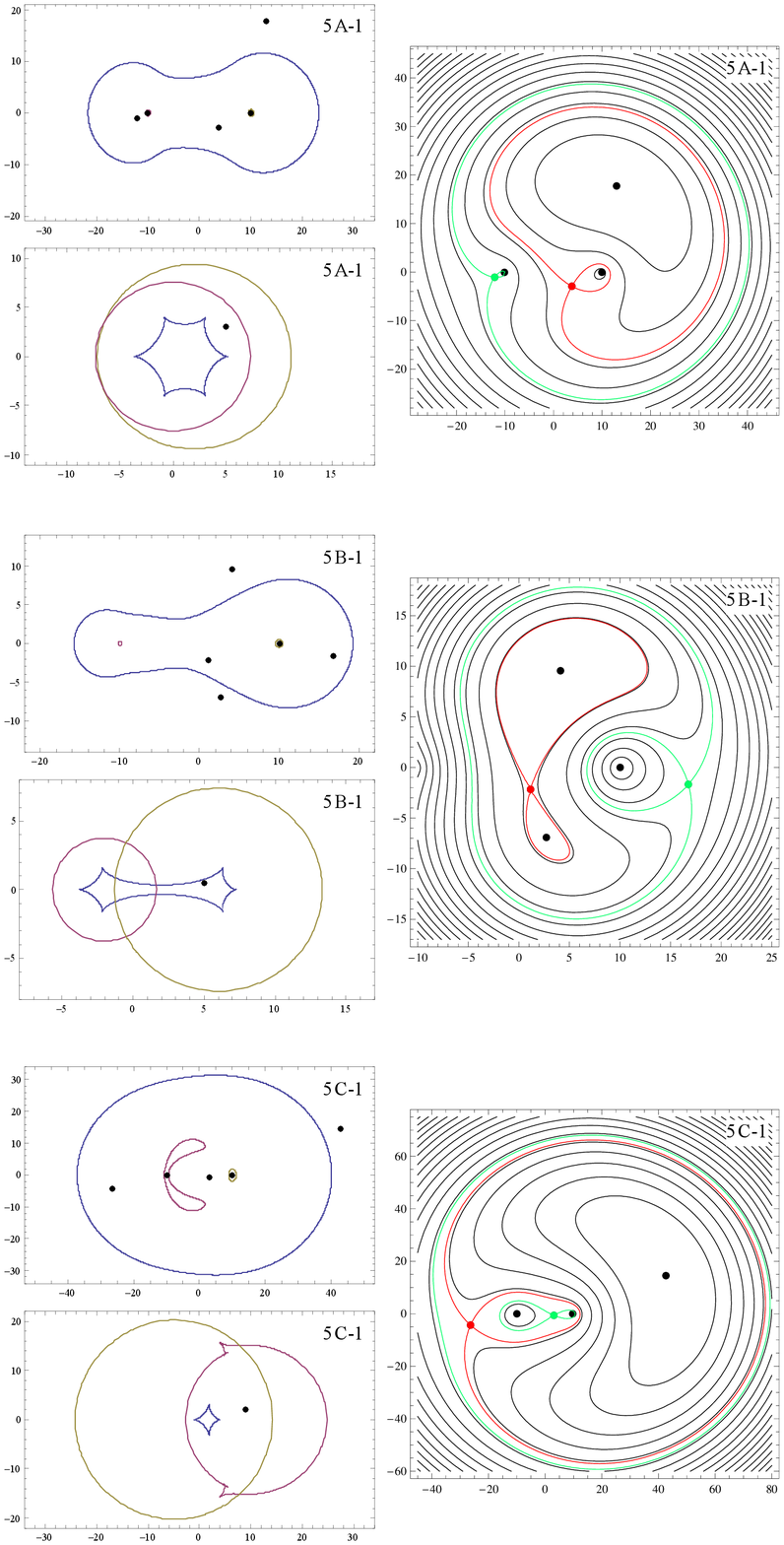}}
\caption{The different five-image Fermat surface topographies and
  image configurations (left- and right-hand panels as for
  Fig.~\ref{fig:BE-BCP-FS_3ims}). The saddle-point images coloured red
  arrive before those coloured green. Examples 5A-1 and 5C-1 would be
  observed as triplets because of the two highly-demagnified images
  near the galaxy centres at $(\pm10,0)$, whilst the others would be
  observed as quadruplets. 5A-2 and 5C-2 are qualitatively different
  image configurations that have the same critical isochrone
  topologies as 5A-1 and 5C-1, respectively. In 5A-2, the source is placed
  inside a three-cusp caustic and one of the radial critical curves,
  so only one of the five images is a highly demagnified core
  image. In 5C-2, a different choice of $\rcOne$ and $\rcTwo$ is made
  so that a swallowtail cusp in the caustic that corresponds to the
  `Pacman' critical curve can pierce the radial caustic. A source
  placed inside the swallowtail but outside the radial caustic gives
  five images, of which only one is highly
  demagnified.  \label{fig:BE-BCP-FS_5ims}}
\end{figure*}

\begin{figure*}
\epsfxsize = 8.4cm 
\flushright{\epsfbox{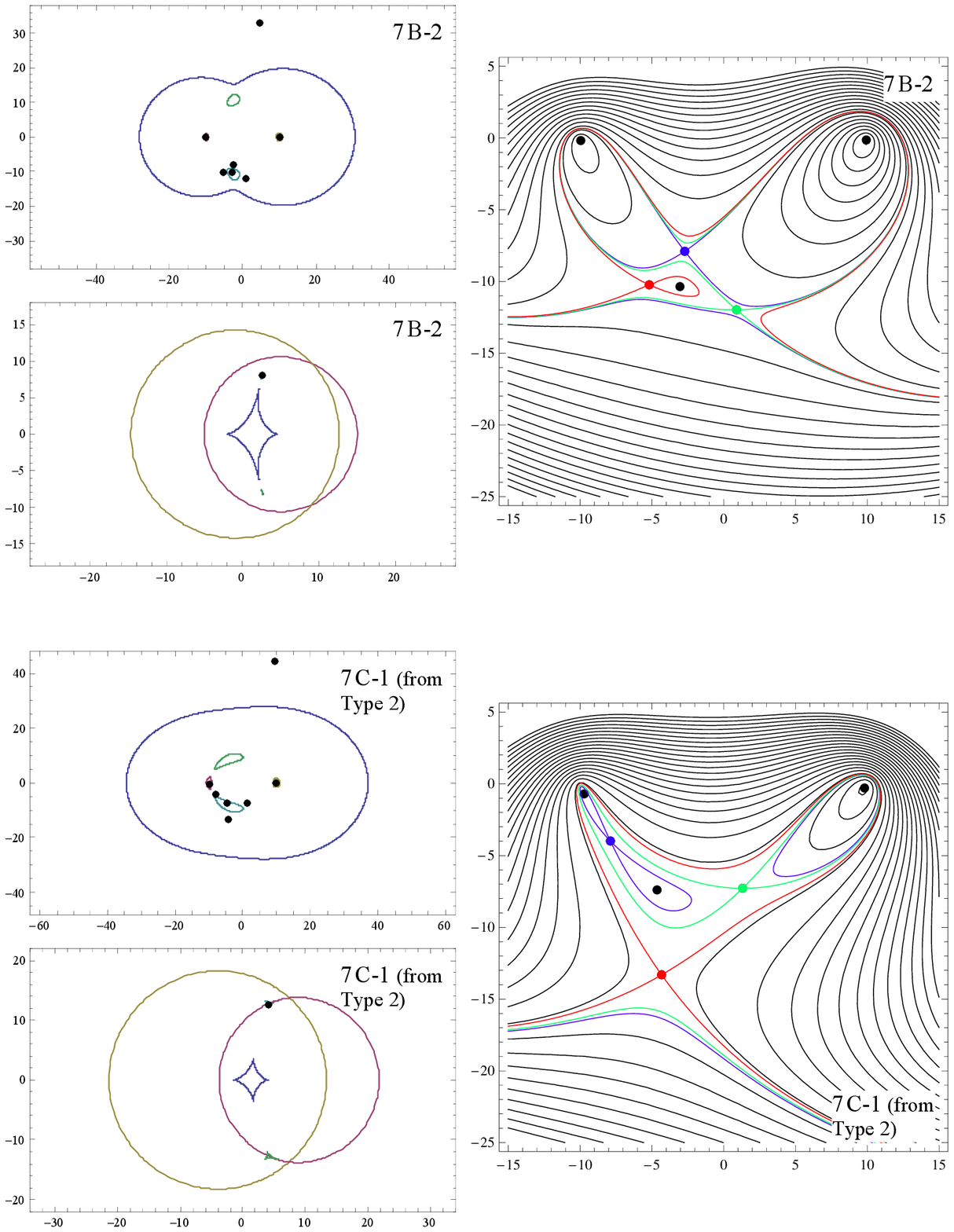}}
\vspace{-13cm}\flushleft{\epsfbox{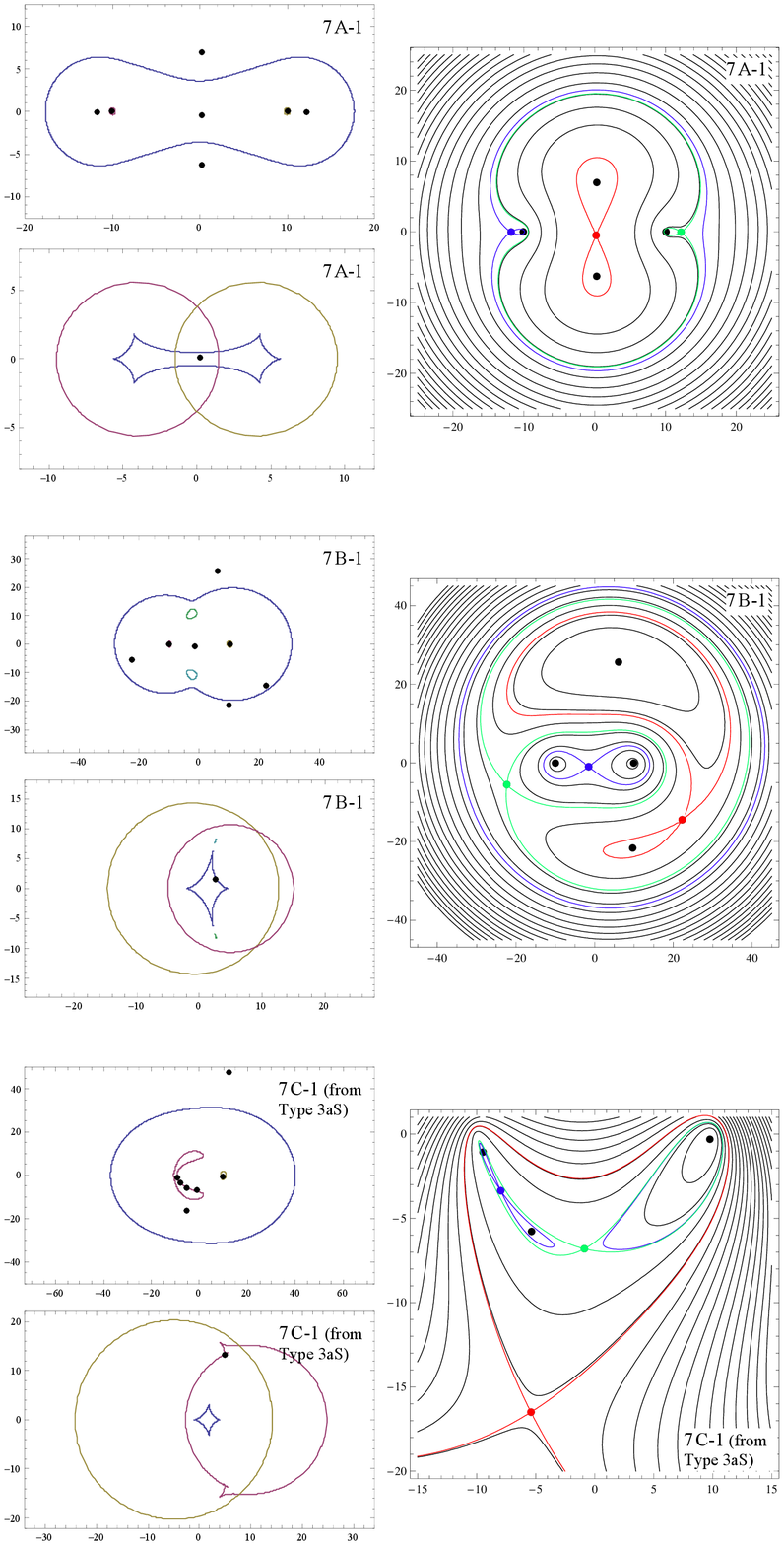}}
\caption{The different seven-image Fermat surface topographies and
  image configurations. The order of arrival of saddle-point images is
  red, green, blue. The two 7C-1 systems are not qualitatively
  different (although the image near $(-10,0)$ is more demagnified),
  but arise from a topologically distinct caustic configurations.
  Except for the 7C-1 from Type 3aS caustics, and possibly the 7C-1
  from Type 2 caustics, these would all be observed as quintuplets,
  as they each have two core images. 7B-2 has the same critical isochrone
  topology as 7B-1, but the image positions are very different and one
  of its critical isochrones (coloured red) is topologically a
  lemniscate but extremely distorted.
  \label{fig:BE-BCP-FS_7ims}}
\end{figure*}

There can be 1, 3, 5 or 7 images; the corresponding regions in the
source plane can be easily deduced in Figures~\ref{fig:BE-BCP-Set-A}
to~\ref{fig:BE-BCP-Metamorphosis-3b}. Only one image is produced if
the source is outside all the caustics, and there are two extra images
for each caustic within which the source lies. Central images are
usually highly demagnified and would not be observable, so some
3-image systems would appear as doublets (e.g. 3A-1 in
Fig.~\ref{fig:BE-BCP-FS_3ims}), some 5-image systems as quadruplets
(e.g. 5B-1 in Fig.~\ref{fig:BE-BCP-FS_5ims}) or even triplets
(e.g. 5A-1 in Fig.~\ref{fig:BE-BCP-FS_5ims}), and 7-image systems
usually as quintuplets (e.g. 7A-1 in
Fig.~\ref{fig:BE-BCP-FS_7ims}). The different image geometries may be
classified according to the topography of the Fermat potential
~\citep[see e.g.,][]{Sc92}, which is
\begin{equation}\label{eq:BE-phi-in-general}
\phi_{\vecxi}(\vecx) = \half(\vecx - \vecxi)^{2} - \psi(\vecx)
\label{eq:fermat}
\end{equation}
where $\vecxi = (\xi,\eta)$ and $\vecx = (x,y)$. A helpful
visualization of eqn. (\ref{eq:fermat}) is provided by the Fermat time
delay surface, examples of which appear below (Figs~\ref{fig:BE-BCP-FS_3ims} to \ref{fig:BE-BCP-FS_7ims}).
The images occur at stationary points of the Fermat surface $\grad\phi
= 0$, which may be local maxima, local minima, or saddles. (We exclude
cases where $y$ lies on a caustic, so there are no images at critical
points.) Now, $\phi(\vecx)$ looks like a concave-up paraboloid at
large $\abs{\vecx}$ because $\psi \to 0$ (or a constant), so level
curves of $\phi(\vecx)$ must be closed. Also, level curves cannot
cross except at saddle points. A Fermat surface can therefore be
classified by the topology of the level curves that run through
saddles, the \textit{critical isochrones}, (see
e.g. \cite{Sc92}~\S5.5). There are four ways to close the critical
isochrones that run through a saddle, as shown in
Fig.~\ref{fig:BE-CritIso-LeLi-diagrams}, which we call Le$^{\pm}$ (the
lemniscates) and Li$^{\pm}$ (the lima\c{c}ons) for short. We recall
that if there are 3, 5 or 7 images, there are 1, 2 or 3 saddle points,
respectively. For a given number of saddle points, it is
straightforward to count the number of possible critical isochrone
topologies: it is a matter of counting the number of different ways in
which lima\c{c}ons and lemniscates can be enclosed within one another
consistently, assuming that $\phi(S_{i}) \neq \phi (S_{j})$ for
different saddle points $S_{i},\,S_{j}$ ~\citep[see
e.g.][]{Sc92}. There are 2 possible topologies for 3-image geometries,
6 for 5-image geometries and 25 for 7-image geometries, but not all of
these actually arise in a given lens model~\citep{Bl86}.

Although there are many different possible topologies for the caustics
(Types 0 to 3b, swallowtails, and different overlapping of caustics),
that divide the source plane up into many different regions, there are
only eight different critical isochrone topologies for the double
isothermal sphere lens: two three-image cases, three five-image cases,
and three seven-image cases. These are listed in
Table~\ref{tab:BE-BCP-FermatSurface-Topographies}, and the possible
five- and seven-image topologies are shown explicitly in
Fig.~\ref{fig:BE-BCP-FSTopologyDiagrams}. We name them according to
the shapes of the critical isochrones, starting from the outermost and
working inwards; the exact placement of the lima\c{c}ons and
lemniscates follows implicitly. So, for example, Li$^{+}$ Li$^{+}$
Le$^{-}$ means that an outer lima\c{c}on encloses an inner lima\c{c}on
(and an Li$^{+}$ can only be put inside the minus region of the outer
lima\c{c}on), which in turn contains an innermost lemniscate (which
must be in the minus region of the inner lima\c{c}on), whilst Li$^{+}$
Le$^{+}$ Le$^{-}$ means that there is a Le$^{+}$ in the plus region of
the lima\c{c}on and a Le$^{-}$ in the minus region (since a Le$^{-}$
cannot be enclosed within a Le$^{+}$).

For the same isochrone topology, image configurations can differ
qualitatively, as listed in the second column of the Table. So, for
example, 7B-1 and 7B-2 are two distinct seven-image configurations
corresponding to Li$^{+}$ Le$^{+}$ Le$^{-}$. Examples of
qualitatively different image configurations are plotted in
Figs~\ref{fig:BE-BCP-FS_3ims} to \ref{fig:BE-BCP-FS_7ims}, with image
positions superimposed on critical curves and Fermat surfaces with
critical isochrones. The galaxies are still always centred at
$(\pm10,0)$ in these figures, and the core radii $r_{c\,1,2}$ are
chosen as before.

Classifying image geometries by critical isochrone topology is of more
than mathematical interest because it allows the order of arrival of
images to be deduced (in whole or in part). But, as the figures show,
systems with the same critical isochrone topology can be very
different: they can have a different number of core images (as in
examples 5C-1 and 5C-2 in Fig.~\ref{fig:BE-BCP-FS_5ims}), or the bright
images may be positioned very differently (e.g. 7B-1 and 7B-2 in
Fig.~\ref{fig:BE-BCP-FS_7ims}), and, since Fermat surfaces change
smoothly with changing lens parameters and source position, two
systems can have similar image positions but have different
critical isochrone topologies. Indeed, more than one critical
isochrone topology is possible for sources in the same region of a
given critical curve and caustic topology: for example, both the
Fermat topographies Li$^{+}$ Li$^{+}$ Le$^{-}$ (e.g. 7A-1) and Li$^{+}$
Le$^{+}$ Le$^{-}$ (e.g. 7B) can arise from a Type-1 caustic
configuration with the source inside all three caustics (not shown in
Figures). It is also possible to obtain the same critical isochrone
topology and similar image positions for lenses with topologically
distinct critical curves and caustics (e.g. the two 7C-1 panels in
Fig.~\ref{fig:BE-BCP-FS_7ims}).

\begin{table*}
\begin{center}
  \caption{The different Fermat surface topographies of a lens consisting of two non-singular isothermal spheres, and the critical
    curve configurations which can give rise to them. \label{tab:BE-BCP-FermatSurface-Topographies}}
\begin{tabular}{llcll}
\hline \\
Number & Critical isochrone   & Name of example    & Critical curve         & Source positions \\
of images & topology             & image configuration & and caustic type      & \null \\
\hline \\
Three images & Li$^{+}$                   & 3A-1    & Any type             & Inside one radial (or `Pacman') caustic only. \\
 & Li$^{+}$                   & 3A-2  & Type 2               & Inside a three-cusp caustic only \\
 & Le$^{-}$                   & 3B-1    & Type 0, Type 1       & Inside astroid or hexacuspid caustic only. \\
& & & &\\
& & & &\\
Five images & Li$^{+}$ Li$^{+}$          & 5A-1 & Type 1, 2            & Inside both radial caustics only. \\
& Li$^{+}$ Li$^{+}$          &               5A-2 & Type 2               & Inside three-cusp caustic and one radial caustic only. \\
& Li$^{+}$ Le$^{-}$          & 5B-1 & Any type             & Inside astroid or hexacuspid and one radial caustic only. \\
& Li$^{+}$ Le$^{+}$          & 5C-1 & Type 3a(S), 3b(S)    & Outside the astroid but inside both other caustics. \\
& Li$^{+}$ Le$^{+}$          &                 5C-2 &Type 3aS             & Inside swallowtail but outside radial caustic. \\
& & & & \\
& & & &\\
Seven images & Li$^{+}$ Li$^{+}$ Le$^{-}$ & 7A-1 & Type 1               & Inside hexacuspid and both radial caustics. \\
& Li$^{+}$ Le$^{+}$ Le$^{-}$ & 7B-1 & Any type & Type 1: inside hexacuspid and both radial caustics. \\
& &             &                            & Type 2: inside both radial caustics and the astroid or a three-cusp caustic. \\
& &             &                            & Types 3a(S), 3b(S): inside astroid and both other caustics. \\
& Li$^{+}$ Le$^{+}$ Le$^{-}$ &  7B-2 & Type 2               & Inside both radial caustics and a three-cusp caustic.\\
& Li$^{+}$ Le$^{+}$ Le$^{+}$ & 7C-1  & Type2                   & Inside both radial caustics and a three-cusp caustic.\\
& &             &                   Types 3aS, 3bS          & Inside swallowtail and the radial caustic.\\
\hline
\end{tabular}
\end{center}
\end{table*}

\section{Two Singular Isothermal Spheres} \label{sec:BE-BSIS}

When the isothermal spheres are singular ($\rcOne = \rcTwo \, = \,0$),
the algebra is simpler than the cored ($\rcOne, \rcTwo \, > 0$) case,
and we derive some analytic results on critical curve and
multiple-imaging regimes. Specifically, we find the pseudocaustics in
\S~\ref{secsub:BE-BSIS-Pseudocaustics}, which allows us (in
\S~\ref{secsub:BE-BSIS-3and5-Imaging}) to derive conditions on the
lens parameters ($E_{1},\,E_{2},\,a$) for which three- and five-fold
imaging becomes possible. We find the lens parameters for
Metamorphosis 1 (in \S~\ref{secsub:BE-BSIS-Met1}), the elliptic
umbilic catastrophe in Type 2 critical curves
(\S~\ref{secsub:BE-BSIS-EUcatastrophe}), and for the singular limit of
Metamorphoses 3a and 3b (\S~\ref{secsub:BE-BSIS-Met3ab}).

\subsection{Pseudocaustics} \label{secsub:BE-BSIS-Pseudocaustics}

The critical curves and caustics can still be of Type 0, 1 and 2 (with
a mild abuse of terminology). The difference between the singular and
cored cases is that the radial critical curves around the two galaxy
centres shrink to zero size as the core radii decrease to zero. This
means that there are no radial caustics, and that any core images move
towards the centres of the isothermal spheres as $\rcOne,\,\rcTwo \to
0$, their magnifications diminishing, until they disappear at the
singularities. A source can therefore produce either an odd or an even
number of images, depending on the number of core images it would
produce if the core radii were not zero. The number of images
changes by one (with an image being created or destroyed at the centre
of a singular isothermal sphere) when the source crosses a
\textit{pseudocaustic} ~\citep[see e.g.][]{Ev98}, a closed curve in
the source plane which is approached by the radial caustics as
$\rcOne,\,\rcTwo \to 0$.

The positions of the pseudocaustics are easily found by considering
the lens mapping on small circles around the singularities. In vector
form, the lens equation reads
\begin{equation}\label{eq:BE-BSIS-LensEqn}
\vecxi = \vecx \:-\: E_{1}\,\frac{\vecx + \vect{a}}{\abs{\vecx +
  \vect{a}}} \:-\: E_{2}\,\frac{\vecx - \vect{a}}{\abs{\vecx - \vect{a}}},
\end{equation}
where $\vect{a} = (a,0)$.  Consider a small circle, say of radius
$\delta$, in the lens plane centred at the second singular isothermal
sphere at $\vect{a}$. Then, as $\delta \to 0$, then $\vecx \to
\vect{a}$ and $(\vecx + \vect{a}) \to (2a,0)$, whilst
$E_{2}\,(\vecx - \vect{a}) / \abs{\vecx - \vect{a}}$ becomes
$E_{2}$ times a unit direction vector. So, the circle around the
singularity at $\vect{a}$ maps to a circle in the source plane centred
at $(a - E_{1},0)$ and with radius $E_{2}$. The other pseudocaustic,
corresponding to the singularity at $-\vect{a}$ is found in the same
way to be circle of radius $E_{1}$, centred at $(-a + E_{2},0)$.

\begin{figure*}
\epsfxsize = 12cm \centreline{\epsfbox{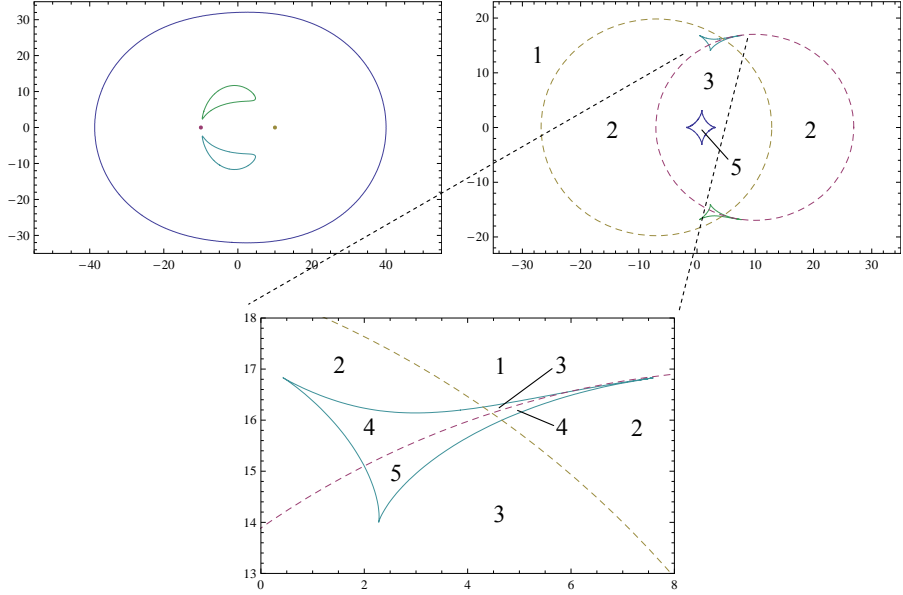}}
\caption{The left panel shows critical curves, and the right panel
  caustics (solid lines), for two singular isothermal spheres at
  $(-10,0)$ and $(+10,0)$ (marked by magenta and ochre dots). The
  pseudocaustics are the dotted circles; when a source crosses a
  pseudocaustic, a single image is created or destroyed at the
  corresponding singularity. The number of images for sources in
  different regions of the source plane are shown. If $\rcOne$ were
  raised from zero, the magenta pseudocaustic would become a true
  caustic as a small radial critical curve formed around $(-10,0)$,
  and if $\rcOne$ were increased further, the three-cusp caustics
  would join with the magenta caustic as a `Pacman' critical curve
  formed (Metamorphosis 3a). Raising $\rcTwo$ from zero would make the
  ochre pseudocaustic a true caustic. \label{fig:BE-BSIS-Example1}}
\end{figure*}

Figure~\ref{fig:BE-BSIS-Example1} shows the critical curve and caustic
configuration for a pair of singular isothermal spheres that, if cores
were put in, would have been of Type 2 and near Metamorphosis 3a. The
number of images produced by point sources in different regions is
shown on the plot of the caustics.  Since the radial critical curves
have shrunk to points, Type 3a and 3b critical curves do not arise in
the case of singular isothermal spheres. Instead, there remain two
three-cusp caustics corresponding to the two small `bean' critical
curves of the Type 2 configuration, although these caustics become
indefinitely elongated along one (both) pseudocaustics for $E_{2} \geq
2a$ (respectively, $E_{1} \geq 2a$), as the `bean' critical curves
then touch the singularities at $(\pm a,0)$ (see
\S~\ref{secsub:BE-BSIS-Met3ab}).

\subsection{The Onset of Three- and Five-Fold imaging} \label{secsub:BE-BSIS-3and5-Imaging}

When the singular isothermal spheres are far apart, so that their
caustics and pseudocaustics do not overlap or join, then a point
source can produce 1, 2 or 4 images. Each isothermal sphere has a
circular pseudocaustic and a small astroid caustic within it, the size
of which shrinks to zero as $a \to \infty$. A source outside the
pseudocaustics produces 1 image, a source inside one of the
pseudocaustics produces 2, and a source inside the astroid produces 4.

As $a$ is decreased, triple imaging becomes possible when the
pseudocaustics overlap.  The condition on the lens parameters for
overlapping pseudocaustics follows from their positions. It is
\begin{equation}\label{eq:BE-BSIS-MultImConds-PseudocausticOverlapCondition}
E_{1} + E_{2} > a.
\end{equation}

\begin{figure*}
\epsfxsize = 15cm \centreline{\epsfbox{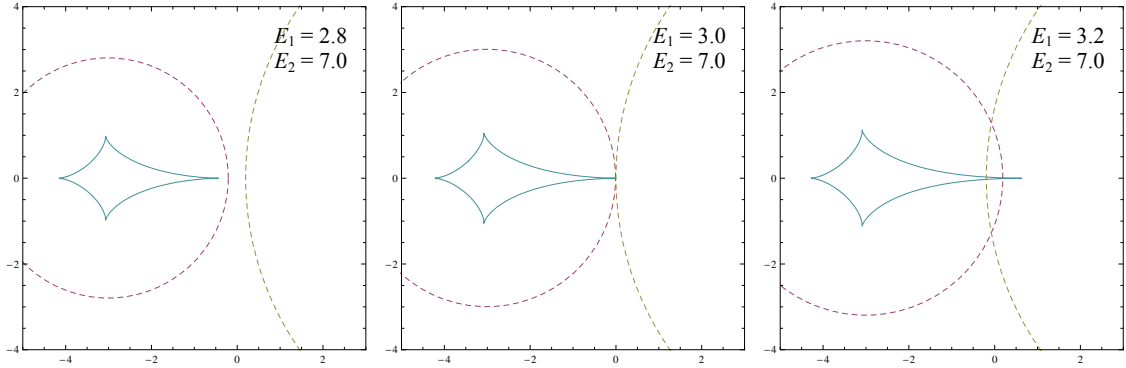}}
\caption{Close-up of the piercing of a pseudocaustic by an astroid
  caustic at the onset of five-fold imaging by a double-SIS
  lens. $E_{i}$ increase from left to right panels. The astroid
  caustics are within their respective pseudocaustics unless the
  pseudocaustics overlap, when the astroid of the less-massive SIS
  (taken w.l.o.g. to be the left-hand one) pierces the pseudocaustic
  of that SIS, and therefore also overlaps with the region enclosed by
  both pseudocaustics. \label{fig:BE-BSIS-5Imaging-Onset}}
\end{figure*}

Five-fold imaging becomes possible once part of an astroid caustic
lies inside both pseudocaustics. The pseudocaustics are given in
\S~\ref{secsub:BE-BSIS-Pseudocaustics}, and it is straightforward to
find the $x$-intercepts of the tangential critical curve(s), and hence
the $\xi$-intercepts of the astroid caustic(s). This allows us to find
the conditions on the lens parameters for which the pseudocaustics
overlap with the astroid. Such overlapping is possible not only for
Type 1 and 2 caustics, but also for Type 0. It turns out that, as $a$
is decreased (or $E_{i}$ increased), five-fold imaging becomes
possible exactly when triple imaging does: when critical curves are of
Type 0, the astroid of the second isothermal sphere stays wholly
within its pseudocaustic (note we still take $E_{1} < E_{2}$, without
loss of generality), whilst the astroid of the first isothermal sphere
pierces its pseudocaustic if and only if the pseudocaustics overlap
(see Fig.~\ref{fig:BE-BSIS-5Imaging-Onset}). We now show this.

The critical curve of the first isothermal sphere cuts the $x$-axis at
two points, and it is the right-hand one -- call it $x_{t1}$ -- that
maps to the right-hand cusp of the astroid that pierces the
pseudocaustic as shown in Fig.~\ref{fig:BE-BSIS-5Imaging-Onset}.
From equations \eqref{eq:BE-Models-kappa} and \eqref{eq:BE-Models-Wyn},
the Jacobian along the $x$-axis for $\rcOne = \rcTwo = 0$ is simple:
\begin{equation}\label{eq:BE-BSIS-detA-xAxis}
\detA(x,0) = 1 - \frac{E_{1}}{\abs{x + a}} - \frac{E_{2}}{\abs{{x-a}}}\:.
\end{equation}
Now, $x_{t1}$ lies between $-a$ and $a$, and the appropriate root of
the quadratic \eqref{eq:BE-BSIS-detA-xAxis} is
\begin{equation}
x_{t1} = \half\left( E_{1} - E_{2} - \sqrt{B^{2} - 4C} \right)\:,
\end{equation}
where $B = E_{2} - E_{1}$ and $C = a(E_{1} - E_{2}) - a^{2}$. The lens
equation gives
\begin{equation}
\xi_{t1} = x_{t1} - E_{1} + E_{2}\,.
\end{equation}
Call the right-hand $\xi$-intercept of the first pseudocaustic
$\xi_{ps1}$. We know from \S~\ref{secsub:BE-BSIS-Pseudocaustics} that
\begin{equation}
\xi_{ps1} = -a + E_{2} + E_{1}.
\end{equation}
Now, the astroid pierces the pseudocaustic if and only if $\xi_{t1} >
\xi_{ps1}$, that is, combining the three equations above, if
\begin{equation}
2a - 3E_{1} - E_{2} - \sqrt{(E_{2}\,-\,E_{1})^{2} + 4a(a\,-\,E_{1}\,-\,E_{2})} \: > \, 0\,,
\end{equation}
which, on squaring and cancelling terms, is equivalent to
\begin{equation}\label{eq:BE-BSIS-5imCond-temp}
(a - 2E_{1})(a - E_{1} - E_{2}) > a(a - E_{1} - E_{2}) \:.
\end{equation}
We are considering critical curves of Type 0, and so -- anticipating
the result \eqref{eq:BE-BCP-Met1-Equation} from the next section
together with the assumption $E_{1} \leq E_{2}$ -- \eqref{eq:BE-BSIS-5imCond-temp} reduces to
\begin{equation}
a - E_{1} - E_{2} < 0\,,
\end{equation}
which is exactly the condition
\eqref{eq:BE-BSIS-MultImConds-PseudocausticOverlapCondition} for
overlapping pseudocaustics. (A similar calculation shows that the
astroid caustic of the second isothermal sphere does not pierce the
second pseudocaustic.)

Calculations for the left- and right-most astroid cusps on the
$\xi$-axis, corresponding to critical points at $x < -a$ and $x > a$,
show that five-fold imaging is always possible if the critical curves
and caustics are of Type 1 or Type 2 (with $E_{1} \leq E_{2} \leq
2a$). However, the astroid may lie entirely outside the first
pseudocaustic if $E_{2} > 2a$. In this case, there is still
overlapping if
\begin{equation}
E_{2}\left(E_{2}-E_{1}\right) < a\left(2E_{2} - E_{1}\right).
\end{equation}

\subsection{The First Metamorphosis} \label{secsub:BE-BSIS-Met1}

By \eqref{eq:BE-Models-detA-Cartesian-psis} and direct differentiation
of the deflection potential \eqref{eq:BE-BCP-psi}, the Jacobian of the
lens mapping for two isothermal spheres reduces, along the $x$-axis,
to
\begin{equation}
\detA(x,0) = R(x) T(x) \:,
\end{equation}
where roots of
\begin{subequations}
\begin{equation}
  R(x) \doteq 1 - {E_{1}\rcOne^{2}\over\left[\rcOne^{2} +
      (x+a)^{2}\right]^{3/2}}-  {E_{2}\rcTwo^{2}\over \left[\rcTwo^{2} + (x-a)^{2}\right]^{3/2}},
\end{equation}
\begin{equation}\label{eqsub:BE-BSIS-TangentialApple}
T(x) \doteq 1 - {E_{1}\over\left[\rcOne^{2} + (x+a)^{2}\right]^{1/2}}
- {E_{2}\over \left[\rcTwo^{2} + (x-a)^{2}\right]^{1/2}},
\end{equation}
\end{subequations}
give radial and tangential critical points, respectively. If $T(x) =
0$ has two roots for $x \in (-a,a)$, then the critical curves are of
Type 0 (there are two separate tangential critical curves). For
singular isothermal spheres, $R(x) \equiv 1$. There are no radial
critical curves, and the three equations above reduce to
\eqref{eq:BE-BSIS-detA-xAxis}, that is,
\begin{displaymath}
\detA(x,0) = T(x) = 1 - \frac{E_{1}}{\abs{x + a}} - \frac{E_{2}}{\abs{{x-a}}}\:.\nonumber
\end{displaymath}
For $x \in (-a,a)$, $T(x)$ is concave down with a single maximum at
some $x_{m}$. $T\dash(x) = 0$ is a quadratic in $x$ with one root in
this interval, and we find
\begin{equation}\label{eq:BE-BSIS-Met1-xm}
x_{m} = -a\:\frac{\sqrt{E_{2}} - \sqrt{E_{1}}}{\sqrt{E_{2}} + \sqrt{E_{1}}}.
\end{equation}
If $T(x_{m}) > 0,$ there are two roots; if $T(x_{m}) < 0,$ there are
no roots. If $T(x_{m}) = 0$, the tangential critical curves are
touching and this is Metamorphosis 1. Combining the two previous
equations, $T(x_{m})$ reads
\begin{equation}\label{eq:BE-BCP-Met1-Equation}
\frac{1}{2a}\:\left( \sqrt{E_1}+\sqrt{E_2} \right)^{2} \: = \: 1\:.
\end{equation}
Critical curves are of Type 0 if the left-hand side (LHS) is less than
$1$ and Type 1 if $\mathrm{LHS} > 1$.  It is clear that
\eqref{eq:BE-BCP-Met1-Equation} remains a very good approximation for
the non-singular isothermal spheres if $\rcOne,\,\rcTwo\: \ll a$.

\subsection{The Elliptic Umbilic Catastrophe} \label{secsub:BE-BSIS-EUcatastrophe}

The radial and tangential critical curves are well-known from single
spherical or elliptical lenses. In such cases, the Jacobian of the
lens mapping, $\detA$, is positive immediately around the galaxy
centre, negative between the radial and tangential critical curves,
and positive outside the tangential critical curve, where $\detA \to
1$ as $\abs{\vecx} \to \infty$. We intuitively expect a double
isothermal sphere lens to have Type 0 critical curves when the
separation $2a$ is large, and for the tangential critical curves to
merge to give Type 1 when the isothermal spheres are brought closer
together. The possibility of Type 2 critical curves, with `bean'
critical curves, may not be as intuitively obvious.

The `bean' critical curves surround small regions of positive $\detA$
inside the common tangential critical curve, where $\detA$ is
otherwise negative. The physical reason that these regions must exist
is that there are points where the shear is zero (c.f. discussion of
the two point-mass lens in \cite{Sc92} \S8.3). From the expression
\eqref{eq:BE-Models-detA-kappa-gamma} for $\detA$, if $\gamma(\vecx) =
0$ at some $\vecx$, then $\detA(\vecx) > 0$.

\begin{figure}
\epsfxsize = 9cm \centreline{\epsfbox{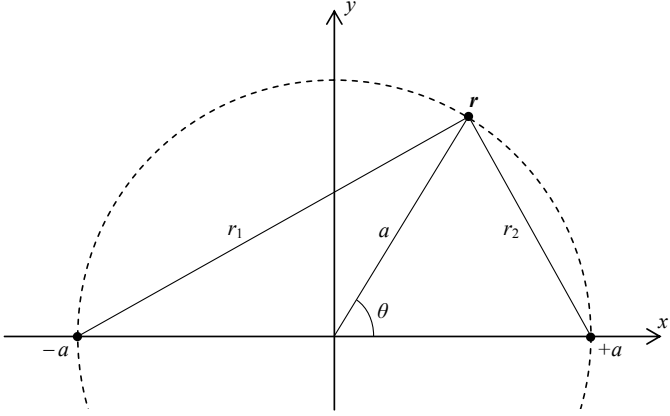}}
\caption{The point of zero shear lies on the circle $\abs{\vecr} =
  a$. The centres of the (singular or cored) isothermal spheres are at
  $\pm a$ on the $x$-axis. \label{fig:BE-Shear-CircleDiagram}}
\end{figure}

For the double isothermal sphere lens, there are indeed two points
where $\gamma = 0$ because the shears from the two isothermal spheres
cancel out.  Let us begin by showing that, if the isothermal spheres
(singular or not) are located at $(\pm a, 0)$, the points of zero
shear lie on a circle of radius $a$ centred at the origin.  It is
useful to define the angle of shear $\omega$ in the usual manner and
write the shear as the complex (polar) quantity
\begin{equation} \label{eq:BE-Shear-Defn-ComplexShear}
\Gamma \doteq \gamma_{1} + i \gamma_{2} \equiv \gamma \, e^{2i\omega}.
\end{equation}
(Note that $\omega \mapsto \omega \pm \pi$ has no physical effect.)
Now, since the deflection potential of two isothermal spheres is the
sum of their two separate deflection potentials, the (complex) shear
is also the sum of two separate shears: $\Gamma = \Gamma^{(1)} +
\Gamma^{(2)}$. Therefore, if the shear is zero at some point $\vecx$,
we must have
\begin{equation} \label{eq:BE-Shear-Omega1-Omega2-Condition}
\omega^{(2)}(\vecx) = \omega^{(1)}(\vecx) + \frac{(2k+1)\pi}{2}\qquad\qquad k \in \Integers.
\end{equation}
For an isothermal sphere, the direction of shear $\omega^{(j)}(\vecx)$ at a
point $\vecx$ must, by symmetry, be either
\begin{equation}
\omega^{(j)}(\vecx) = \theta_{j} \quad\text{or}\quad \omega(\vecx) = \theta_{j} - \frac{\pi}{2}\:,
\end{equation}
where $\theta_{j}$ is the direction of $\vecx$ from the centre of the $j$th
isothermal sphere. (It turns out to be the latter.) Therefore,
\eqref{eq:BE-Shear-Omega1-Omega2-Condition} holds if
\begin{equation}
\theta_{2} - \theta_{1} = \pm\frac{\pi}{2}\:,
\end{equation}
that is, if the line segment joining $(-a,0)$ to $\vecx$ is
perpendicular to that joining $(a,0)$ to $\vecx$. The locus of such
points $\vecx$ is the circle with the two galaxy centres as its
diameter, $\abs{\vect{r}} = a$ (see Fig.~\ref{fig:BE-Shear-CircleDiagram}).

It remains to show that there must exist two points on this circle for
which $\gamma^{(1)} = \gamma^{(2)}$. For a single isothermal sphere,
the magnitude of shear is
\begin{equation}\label{eq:BE-Shear-SingleIS-gamma}
\gamma^{(j)} = \half E_{j} \frac{r_{j}^{2}}{(\rc^{2} + r_{j}^{2})^{(3/2)}},
\end{equation}
where $r_{j}$ is the distance from its centre. As we are only considering
$ \rcOne,\rcTwo \: \ll \: a, E_{1}, E_{2}\:,\;$ at points on the
circle close enough to the first isothermal sphere (the one at
$(-a,0)$), we have $\gamma^{(1)} > \gamma^{(2)}$, whilst the reverse
inequality holds at points close enough to the second isothermal
sphere. There are therefore two points in-between (one in the $y > 0$
semicircle and the other in the $y < 0$) where equality holds.

Although this result does not allow the conditions for Metamorphosis 2
to be determined analytically, it does allow us to find the lens
parameters at which the elliptic umbilic catastrophe occurs: it occurs
when the point of zero shear is also a critical point. We first need
to know the (polar) coordinates $(a,\theta_{zs})$ of the point of zero
shear. Recalling that $r_{1}$ and $r_{2}$ are the distances between a
point $\vecr$ and the centres of the isothermal spheres, on the circle
$\abs{\vecr} = a$ we have
\begin{equation}
\label{eq:BE-Shear-r1Andr2-OnCircle-From-polars}
r_{1}^{2} = 2 a^{2} (1 + \cos\theta), \qquad\qquad
r_{2}^{2} = 2 a^{2} (1 - \cos\theta)\,
\end{equation}
by the cosine rule (see Fig.~\ref{fig:BE-Shear-CircleDiagram}).
Now, by \eqref{eq:BE-Shear-SingleIS-gamma}, the point of zero shear
for the double lens has
\begin{equation}
\frac{r_{1}}{r_{2}} = \frac{E_{1}}{E_{2}},
\end{equation}
and combining this with
\eqref{eq:BE-Shear-r1Andr2-OnCircle-From-polars} yields the result
\begin{equation}\label{eq:BE-Shear-Costheta-of-ZeroShearPoint}
\cos\theta_{zs} = \frac{E_{1}^{2} - E_{2}^{2}}{E_{1}^{2} + E_{2}^{2}}.
\end{equation}

An elliptic umbilic catastrophe occurs at the point of zero shear
(where the surrounding critical curves -- and the corresponding
three-cusp caustics -- shrink to zero size) when
\begin{equation}\label{eq:BE-BSIS-EUCat-Condition}
a = \half \sqrt{E_{1}^{2} + E_{2}^{2}} \:.
\end{equation}
This is proved as follows. From \eqref{eq:BE-Models-detA-kappa-gamma},
at the point of zero shear,
\begin{equation}
\detA = (1-\kappa)^{2}
\end{equation}
where the convergence $\kappa$ is
\begin{equation}\label{eq:BE-Shear-kappa-as-Function-of-r1-r2}
\kappa(r,\theta) = \kappa^{(1)} + \kappa^{(2)} = \frac{E_{1}}{2 r_{1}} + \frac{E_{2}}{2 r_{2}}.
\end{equation}
Now, $r_{1}$ and $r_{2}$ at the point of zero shear are given in terms
of $E_{1}, E_{2}$ and $a$ by
\eqref{eq:BE-Shear-r1Andr2-OnCircle-From-polars} and
\eqref{eq:BE-Shear-Costheta-of-ZeroShearPoint}, and putting
\eqref{eq:BE-Shear-kappa-as-Function-of-r1-r2} into $(1 - \kappa)^{2}
= 0$ yields, after some algebra, \eqref{eq:BE-BSIS-EUCat-Condition}.

\subsection{The Singular Limit of Metamorphoses 3a and
  b}\label{secsub:BE-BSIS-Met3ab}

\begin{figure}
\epsfxsize = 9cm \centreline{\epsfbox{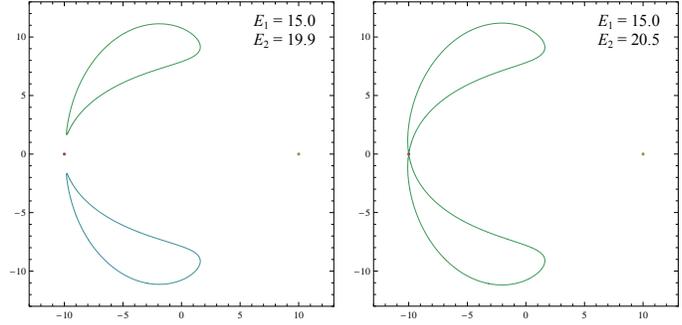}}
\caption{The inner critical curves of two singular isothermal spheres for $a = 10$. The `bean' critical curves meet at the singularity when $E_{2}$ reaches $2a = 20$. \label{fig:BE-BSIS-Met3a-equivalent}}
\end{figure}

As $E_{2} \to 2a$ or $E_{1} \to 2a$ from below, the `bean' critical
curves (that surround points of zero shear) approach the singularity
at $(-a,0)$ or $(a,0)$, respectively, as illustrated in
Fig.~\ref{fig:BE-BSIS-Met3a-equivalent}.
This can be seen from two observations.

First, near the singularities at $(\pm a,0)$, $\detA < 0$ on the
x-axis by inspection of \eqref{eq:BE-BSIS-detA-xAxis}.  Secondly,
along the circle $\abs{\vecr} = a$, $\detA < 0$ near $(a,0)$ or
$(-a,0)$ if and only if $E_{1} < 2a$ or $E_{2} < 2a$,
respectively. This follows from inspection of
\begin{eqnarray}
\detA(a,\theta) &=& \frac{1}{2 a^{2}\, \sin\theta} \Bigl[ E_{1} E_{2}
  + 2 a^{2}\abs{\sin\theta} \nonumber \\
& - & 2 a \left( E_{1}\abs{\sin(\theta/2)} + E_{2}\abs{\cos(\theta/2)} \right) \Bigr] \:,
\end{eqnarray}
which can be obtained -- a little tediously but straightforwardly --
either from
\begin{equation}
\detA(r,\theta) = \frac{1}{r} \left| \begin{array}{cc}
                                             \partial\xi / \partial r & \partial\xi / \partial\theta \\
                                             \partial\eta / \partial r & \partial\eta / \partial\theta \\
                                           \end{array} \right|
\end{equation}
(with the derivatives being evaluated directly from the lens equation
in polar coordinates), or from \eqref{eq:BE-Models-detA-kappa-gamma},
\eqref{eq:BE-Shear-r1Andr2-OnCircle-From-polars},
\eqref{eq:BE-Shear-kappa-as-Function-of-r1-r2} and
\begin{equation}
\gamma(a,\theta) = \abs{\gamma^{(1)} - \gamma^{(2)}} = \half\abs{\frac{E_{1}}{r_{1}} - \frac{E_{2}}{r_{2}}} \:.
\end{equation}

\section{External Shear or Internal Ellipticity}\label{sec:BE-ExternalShear-or-Ellipticity}

In this section, we investigate two astrophysically important
effects. First, Binary galaxies may be embedded in a loose group or
cluster (as in fact is the case for CASSOWARY 2), in which case tidal
shear will distort the caustics and critical curves. Second, galaxies
are characteristically flattened, and so internal ellipticity will
also play a role. In this section, we introduce simple models in
\S~\ref{secsub:shear}, survey the principal effects in
\S~\ref{secsub:effects} and give some analytic results in
\S~\ref{secsub:analytic}

\subsection{Models}
\label{secsub:shear}

\subsubsection{External Shear}

The effects of a larger-scale gravitational field, arising from the
environment in which the lens sits, are often approximated by
external shear: the quadratic terms of the large-scale field
expanded about the centre of the lens. With suitable translation of
the source-plane coordinates, the lens equation including external
shear is
\begin{equation}\label{eq:BE-ExtShear-LensEqn}
\left( \begin{array}{c}
         \xi  \\
         \eta
       \end{array} \right) = \left( \begin{array}{c}
                                      x \\
                                      y
                                    \end{array} \right)  -  \left( \begin{array}{c}
                                                                     \psi_{x}(x,y) \\
                                                                     \psi_{y}(x,y)
                                                                   \end{array} \right)  -  P \left(\begin{array}{c}
                                                                                                     x \\
                                                                                                     y
                                                                                                   \end{array}\right)
\end{equation}
where $\psi$ is the deflection potential of the two isothermal spheres and
\begin{equation}
P \: = \: \left(\begin{array}{cc}
                  \gamma\cos 2\phi & \gamma\sin 2\phi \\
                  \gamma\sin 2\phi & -\gamma\cos 2\phi
                \end{array}\right)
\end{equation}
is the external shear matrix, and $\gamma$ and $\phi$ are the
magnitude and angle of external shear respectively. (Note that $\phi =
0$ corresponds to shear from a perturbing mass at $\infty$ or
$-\infty$ on the $y$-axis). The Jacobian of the lens mapping becomes
\begin{equation}\label{eq:BE-ExtShear-detA}
\detA = \left( 1 - \psixx - P_{11} \right) \left( 1 - \psiyy - P_{22} \right) - \left( \psixy + P_{12} \right)^{2} \:.
\end{equation}
By the arguments of \S~\ref{sec:BE-BSIS}, the pseudocaustics are still
circles of radii $E_{1}$ and $E_{2}$, but are now centred at
\begin{subequations}
\begin{equation}
\left( -a + E_{2} + \gamma a \cos 2\phi \: , \: \gamma a \sin 2\phi \right)
\end{equation}
and
\begin{equation}
\left( a - E_{1} - \gamma a \cos 2 \phi  \: , \: - \gamma a \sin 2\phi \right) \:.
\end{equation}
\end{subequations}

\subsubsection{Ellipticity}

Galaxies are also characteristically flattened. One commonly used
flattened generalization of the isothermal sphere is the
pseudo-isothermal elliptic potential
\citep[see.e.g.][]{KK93,Hu01,Ev02}, which coincides with the
isothermal sphere model when its ellipticity parameter $\epsilon$ is
zero, and otherwise has elliptical isopotentials with axis ratio
$(1+\epsilon)/(1-\epsilon)$. The models are the projections of the
power-law galaxies studied by~\citet{Ev94}.  When the isothermal
spheres at $(\pm a, 0)$ are replaced with this model, the deflection
potential of the lens becomes
\begin{equation}
\psi(x,y) = {E_{1}\over (r_{c,1}^2 + r_1^2)^{1/2}} + {E_{2}\over (r_{c,2}^2 + r_2^2)^{1/2}}
\end{equation}
where
\begin{eqnarray}
r_1^2 & = & x'^2 (1-\epsilon) + y'^2(1+ \epsilon)\\\nonumber
r_2^2 & = & x''^2 (1-\epsilon) + y''^2(1+ \epsilon)
\end{eqnarray}
and $(x',y')$ and $(x'',y'')$ are Cartesians centred on $(-a,0)$ and
$(a,0)$ inclined at $\phi_{1}$ and $\phi_2$ to the $x$-axis
respectively.

\subsection{Effects on Critical Curves, Caustics and Image Multiplicity}
\label{secsub:effects}

\begin{figure*}
\epsfxsize = 14.5cm \centreline{\epsfbox{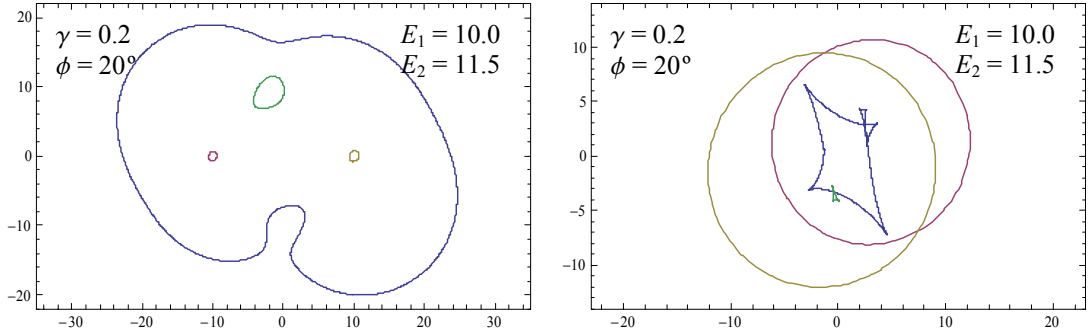}}
\caption{The critical curves (left panel) and caustics (right panel)
  of two isothermal spheres in moderately strong external shear. With
  shear not along the $x$ or $y$ axes, Metamorphosis 2, the
  splitting-off of two `bean' critical curves from the common
  tangential critical curve, is split into two separate
  metamorphoses as one `bean' is created before the
  other. \label{fig:BE-ExtShear-BCP-Met-2-2-example}}
\end{figure*}

\begin{figure*}
\epsfxsize = 14.5cm \centreline{\epsfbox{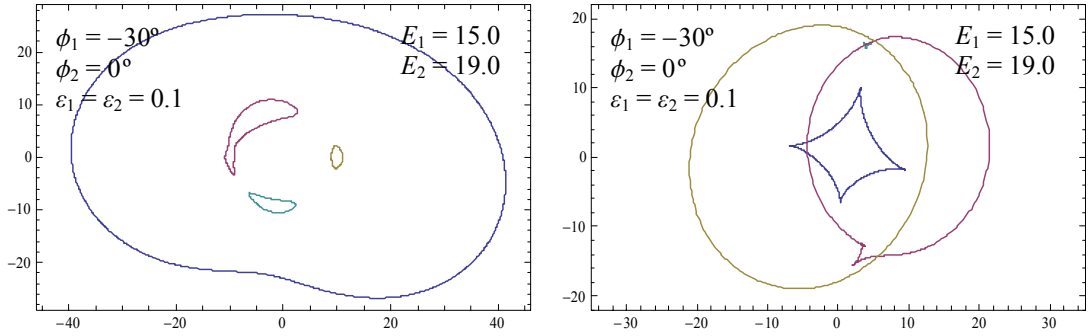}}
\caption{The critical curves (left panel) and caustics (right panel)
  of two pseudo-isothermal elliptical potentials both with $\epsilon =
  0.1$. Unless the axes of the elliptical potentials are both aligned
  with the $x$ or $y$ axes, Metamorphosis 3a is split into two
  separate metamorphoses as one `bean' critical curve merges with a
  radial critical curve before the other does. Metamorphosis 3b is
  similarly split.\label{fig:BE-ExtShear-BPIEP-Met-3a-2-example}}
\end{figure*}

A numerical investigation of the critical curves and caustics
illustrates the essential differences between these models and the two
isothermal spheres.  For realistic values of shear or ellipticity, no
dramatically different critical curve topologies arise, although there
are minor differences resulting from the breaking of the reflection
symmetry about the $x$-axis: Metamorphoses 2, 3a, and 3b can occur in
two steps, with one `bean' critical curve being created, or merging
with a radial critical curve, before the other one -- examples are
presented in the panels of
Figs.~\ref{fig:BE-ExtShear-BCP-Met-2-2-example} and
\ref{fig:BE-ExtShear-BPIEP-Met-3a-2-example}.

More interestingly perhaps, the maximum number of images can change
with the addition of shear or ellipticity. Of course, when shear or
ellipticity is added to a single, isolated isothermal sphere, the
maximum possible number of images increases from three to five (or
from two to four in the singular case) because the tangential caustic,
which had been a single point, becomes an astroid caustic with the
breaking of circular symmetry. There is not always such a dramatic
difference when shear or ellipticity is added to a double isothermal
sphere lens, because circular symmetry has already been
broken. However, the tangential critical curve(s) is (are) distorted
by shear (see, for example, the left panel of
Fig.~\ref{fig:BE-ExtShear-BSIS-Type-0-example}). Shear with $\phi
\approx 0\degrees$, or ellipticity with $\phi_{i} \approx 90\degrees$,
can lead to extra cusps forming in the tangential caustic(s), which
can then self-overlap. When there are two separate tangential
caustics, this can increase the maximum number of images from five to seven
(from four to six in the singular case); when there is a common tangential
caustic, there can be regions of nine- and even eleven-fold imaging (seven- or
nine-fold for the singular case).

\begin{figure*}
\epsfxsize = 14.5cm \centreline{\epsfbox{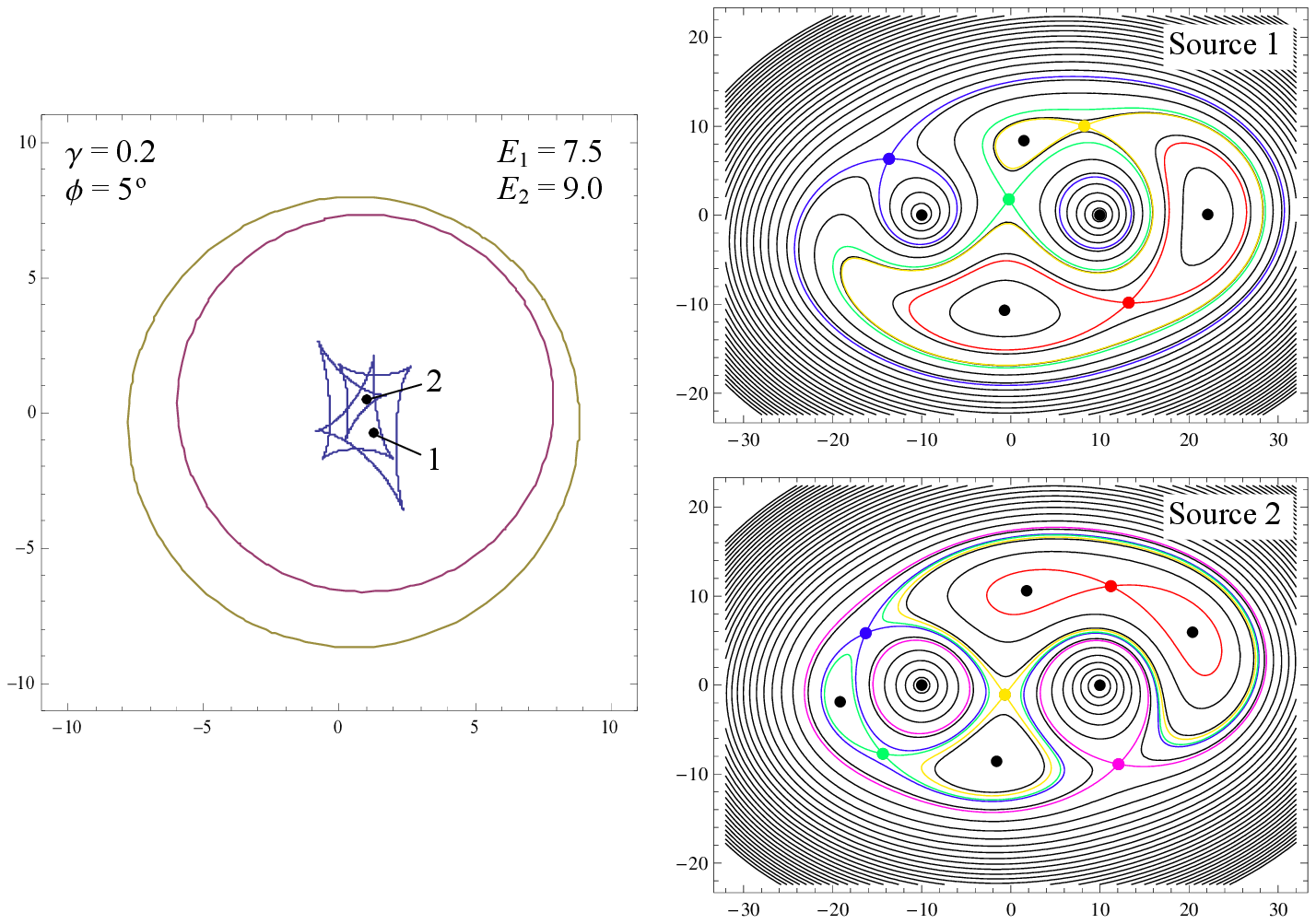}}
\caption{The lens here is two cored isothermal spheres at $(\pm a, 0)$
  with external shear of $\gamma = 0.2$ in the direction $ \phi =
  5\degrees $. Without shear the critical curves and caustics would be
  Type 1. The positions of two point sources are shown with the
  caustics in the left panel; the right panels are contours of the
  Fermat potential for the two sources, together with saddle-point
  images and their critical isochrones (coloured points and lines) and
  the images (black points) corresponding to Fermat maxima and
  minima. The order of arrival of the saddle-point images is red,
  yellow, green, blue, magenta. The four saddles of the nine-image
  case give a critical isochrone topology Li$^{+}$ Li$^{+}$ Le$^{-}$
  Le$^{-}$, and two of the images are probably unobservable `core'
  images; the five saddles of eleven-image case give Li$^{+}$ Li$^{+}$
  Le$^{-}$ Le$^{-}$ Le$^{-}$ and nine of the images are
  observable. \label{fig:BE-ExtShear-BCP-Type1-HighOrderIm}}
\end{figure*}

The left panel of Fig.~\ref{fig:BE-ExtShear-BCP-Type1-HighOrderIm}
shows the caustics for two cored isothermal spheres in moderately
strong external shear aligned close to the $x$-axis. The originally
six-cusped caustic develops four extra cusps, and the self-overlapping
creates regions of high-order imaging: Source 1 is imaged nine times, and
Source 2 imaged eleven times. The right-hand panels of
Fig.\ref{fig:BE-ExtShear-BCP-Type1-HighOrderIm} show the Fermat
surfaces for the two sources, along with image positions and the
critical isochrones. Note that the elevenfold imaging configuration
bears a remarkable similarity to the recently discovered lens
CASSOWARY 2, although of course this is not a fit.

\subsection{Analytic Results}\label{secsub:analytic}

One simple result remains analytic for a double isothermal sphere lens
in external shear: the touching of the two critical curves along the
$x$-axis.  When the isothermal spheres are far apart, the shear
distorts the tangential caustics into ellipses with minor axes aligned
with the direction of external shear, as expected from the effect of
shear on one isothermal sphere \citep[see e.g.][]{Saas-Fee}. The
tangential caustics are of course no longer ellipses when the
isothermal spheres are brought closer together, but they still tend to
be stretched in the direction perpendicular to the external shear. We
expect shear with $\phi = 0\degrees$ ($90\degrees$) to shift
Metamorphosis 1 to larger (smaller) $E_{1}, \,E_{2}$ compared to the
shearless case. This is indeed the case.

\begin{figure}
\epsfxsize = 9cm \centreline{\epsfbox{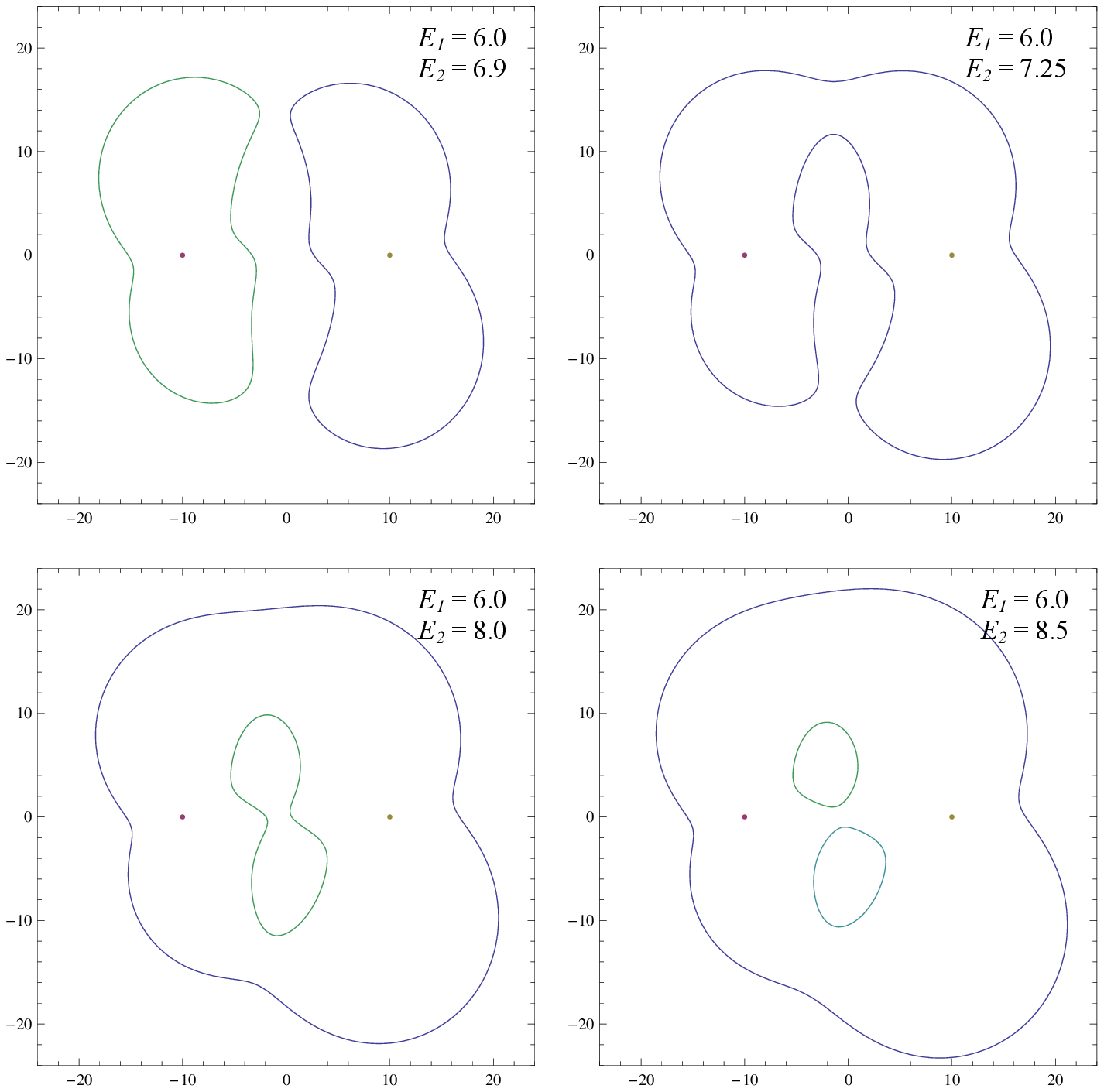}}
\caption{Critical curves of lenses with two singular isothermal
  spheres (dots at $(\pm a,0)$) in very strong external shear. In all
  four panels, the external shear is $\gamma = 0.5$, $\phi =
  10\degrees$. The strong shear distorts the tangential caustics of
  the singular isothermal spheres. In particular, it prevents them
  merging on the $x$-axis until $E_{i}$ are much larger: without
  external shear, the lenses in all four panels would have a single
  common tangential caustic. The condition
  \eqref{eq:BE-ExtShear-Met1-Equation} holds at the transition between
  the topologies of the bottom-left and bottom-right
  panels. \label{fig:BE-ExtShear-BSIS-Met-A}}
\end{figure}

\begin{figure}
\epsfxsize = 9cm \centreline{\epsfbox{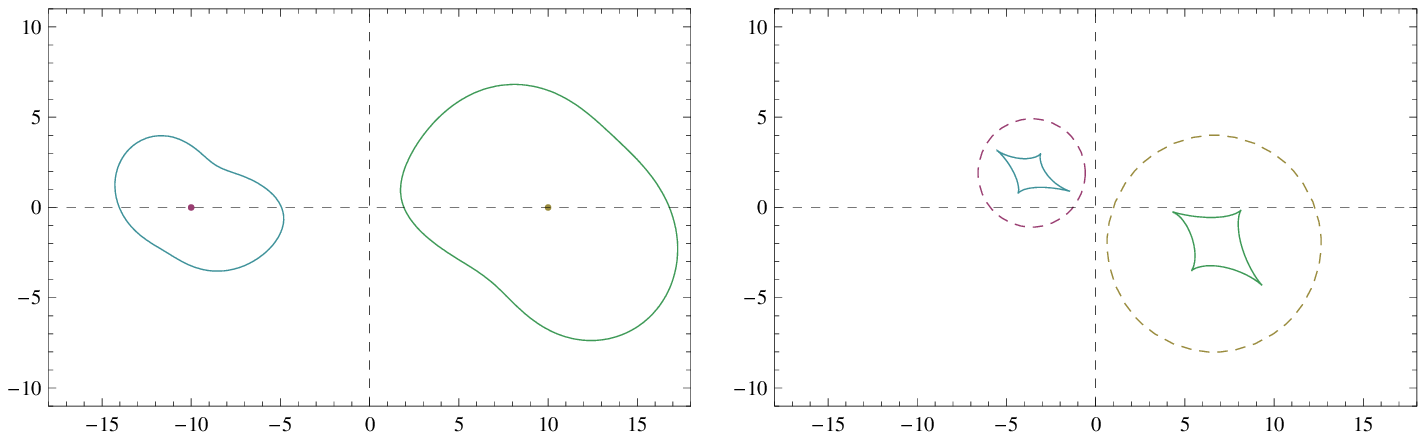}}
\caption{The critical curves (left panel) and caustics (right panel)
  of a double singular isothermal sphere lens with $a = 10$, $E_{1} =
  3.0,$ $E_{2} = 6.0$ and moderate external shear ($\gamma = 0.2$) in
  the direction $\phi = 40\degrees$, which distorts the tangential
  critical curves (and caustics). It is believable, but not obvious,
  that the critical curves will meet on the $x$-axis if $E_{i}$ are
  increased. \label{fig:BE-ExtShear-BSIS-Type-0-example}}
\end{figure}

There is a complication, however. Since even a small nonzero $\gamma$ at
$\phi \neq 0\degrees,\,90\degrees$ distorts the critical curves (see
Figs.~\ref{fig:BE-ExtShear-BSIS-Met-A} and
\ref{fig:BE-ExtShear-BSIS-Type-0-example}) and destroys the reflection
symmetry in the $x$-axis, it is no longer clear that $\detA(x,y)$ must
have a saddle point on the $x$-axis as it did when there was no
external shear. Indeed, the two tangential critical curves do not
necessarily first merge at the $x$-axis: for very strong shear
($\gamma \gtrsim 0.45$ for $a = 10$) and small $\phi$, the tangential
critical curves may merge at two points above and below the $x$-axis
first, giving rise to new critical curve geometries, examples of which
are shown in the panels of
Fig.~\ref{fig:BE-ExtShear-BSIS-Met-A}. These new critical curve
metamorphoses don't occur if there is internal ellipticity but no
shear.


However, as we now show, in the case of singular isothermal spheres,
there is still always a saddle point in $\detA(x,y)$ on the $x$-axis,
which, for reasonable values of $\gamma$, \textit{is} the point at
which the tangential critical curves first touch (when $\detA = 0$ at
the saddle point). First, we obtain $\del\detA$ from
\eqref{eq:BE-ExtShear-detA} and direct differentiation. Restricted to
the $x$-axis, we have
\begin{subequations}
\begin{equation}
\pdiff{x}\detA = \left( \frac{E_{1} (x+a)}{\abs{x+a}^{3}} \!+\! \frac{E_{2} (x-a)}{\abs{x-a}^{3}} \right) \: \left( 1\!-\!\gamma \cos 2\phi \right) \,,
\end{equation}
\begin{equation}
\pdiff{y}\detA = \left( \frac{E_{1} (x+a)}{\abs{x+a}^{3}} \!+\! \frac{E_{2} (x-a)}{\abs{x-a}^{3}} \right) \: 2 \gamma \sin 2\phi \,,
\end{equation}
\end{subequations}
so it is clear that there is one stationary point on the x-axis for
$x\in (-a,a)$ because $\pdiff{x}\detA|_{y=0}$ and
$\pdiff{y}\detA|_{y=0}$ have a common root. This stationary point is a
maximum of
%
%
\begin{eqnarray}
\detA(x,0) &=& \left( 1 + \gamma \cos 2\phi - \frac{E_{1}}{\abs{x+a}} - \frac{E_{2}}{\abs{x-a}} \right)\nonumber\\
&\times& \left( 1 - \gamma \cos 2\phi \right)  - \gamma^{2}\sin^{2}2\phi \:,
\end{eqnarray}
which for $\gamma = 0$ reduces to \eqref{eq:BE-BSIS-detA-xAxis}. By
the argument of \S~\ref{secsub:BE-BSIS-Met1}, the tangential critical
curves touch each other on the $x$-axis when
\begin{eqnarray}\label{eq:BE-ExtShear-Met1-Equation}
  \left( 1 - \gamma \cos 2\phi \right) \left( 1 + \gamma \cos 2\phi -
    \frac{\left(\sqrt{E_{1}} + \sqrt{E_{2}}\right)^{2}}{2a} \right) &  &
  \nonumber \\
  = \gamma^{2}\sin^{2}2\phi & & \:.
\end{eqnarray}
The condition on the lens parameters for this metamorphosis is the
analogue in the unsheared case of \eqref{eq:BE-BCP-Met1-Equation}.
There is always a saddle on the $x$-axis, no matter how large $\gamma$
is. However, the merger that occurs at the saddle on the $x$-axis may
not be the first such merger between the two tangential critical
curves if $\gamma$ is large, as illustrated in the panels of
Fig.~\ref{fig:BE-ExtShear-BSIS-Met-A}.
Eqn~\eqref{eq:BE-ExtShear-Met1-Equation} gives the condition for the
bottom-left panel changing to the bottom-right.  Once again, this
criterion holds approximately if $\rcOne,\rcTwo$ are small but
nonzero.


\section{Conclusions}

This paper has provided a detailed investigation into the lensing
properties of binary galaxies, modelled as two isothermal spheres with
different velocity dispersions.  Such models are useful for describing
strong lenses in which two close galaxies bend light rays emanating
from a more distant source. With the addition of external shear, the
models can also represent groups or clusters dominated by a pair of
massive galaxies. This is of course a very common circumstance, with
about a quarter of all strong lenses showing evidence for lensing by
pairs or groups of galaxies.

Our paper extends and generalizes the previous binary lens work of
\citet{Sc86}, who used two point masses. Binary isothermal spheres
differ qualitatively from the better-known case of two point masses by
having additional critical curves and caustics (or pseudocaustics in
the case of singular isothermal spheres), and therefore also different
multiple-imaging properties. If the isothermal spheres have cores,
then the number of images can be 1, 3, 5 or 7, depending on the
location of the source, the separation of the binary and the velocity
dispersions. If the isothermal spheres are coreless, then the number
of images may be 1, 2, 3, 4 or 5. This lens model is sufficiently
simple that many of the properties can be derived analytically, or
nearly so. We have described the shapes of the caustics and critical
curves for the case of equal and unequal masses, have studied the
principal metamorphoses and have classified the critical isochrone
topologies. Finally, we presented a brief survey of the effects of the
inclusion of external shear and internal flattening.

For comparison with the real world, it is helpful for us to show how
our models relate to physical systems.  The typical velocity
dispersion of a large elliptical galaxy is $\sigma \sim 250$-$300
\kmpersec$, while typical core radii are in the range $\sim
10$-$100\pc$. Given a typical lens redshift of $\zl \sim 0.5$ and
source redshift of $\zs \sim 2$, then the corresponding Einstein
radius of an isolated isothermal sphere is $\sim 10$ kpc. The effects
studied in this paper become important if two galaxies are separated
by $\lesssim$ 4 Einstein radii (see e.g., Fig~\ref{fig:BE-BCP-Set-A}).
At the typical redshifts, this corresponds to a separation of
$\lesssim 40$ kpc (or $\lesssim 5^{\prime\prime}$). A good example is
provided by CASSOWARY 5, which has two lensing galaxies separated by
$\sim 5^{\prime\prime}$~\citep{Be08}. The three images straddle the two
lensing galaxies, much as the illustration in the lower panel of
Fig.~\ref{fig:BE-BCP-FS_3ims}. CASSOWARY 5 looks like an example of
our image configuration 3B-1, though it important to bear in mind that
the existing, primarily Sloan Digital Sky Survey, data is
comparatively shallow ($r \lesssim 23$).  Deeper follow-up imaging may
revel the presence of more images in CASSOWARY 5.

There are some interesting ways to extend the work in this
paper. First, the effects of nearby galaxies are often modelled with
external shear, so a thorough comparison of the properties of binary
isothermal lenses with the single isothermal lens plus shear is
warranted. As is well-known, the Chang-Refsdal lens (or point mass
plus external shear) is only partially successful in describing the
lensing properties of binary point masses. Second, although we have
sketched the principal effects of flattening and shear here, it would
be interesting to prove the maximum number of possible images. The
fact that caustics can have points of self-intersection means that
regions of the source plane are created for which very high-order
imaging is possible. Third, there are still very few highly
non-axisymmetric lenses that have been studied. The binary lens
(whether point masses or double galaxies) is one example, but it would
be interesting to study the lensing properties of needles, filaments,
and sheets, as well. These may have applications to highly prolate or
highly flattened galaxies, which are the natural endpoints of the
Newtonian collapse of spheroids of collisionless matter~\citep{Li65}.

\section*{Acknowledgements}
EMS acknowledges financial support from the Commonwealth Scholarship
Commission and Cambridge Commonwealth Trust. This research has been
supported by the `ANGLES' European Union Research and Training Network
for gravitational lensing. We thank an anonymous referee for a careful
reading of the manuscript.



\label{lastpage}

\end{document}